\documentclass[aps,prd,superscriptaddress,nofootinbib,showpacs,twoside,notitlepage,reprint]{revtex4-1}

\usepackage{amsmath}
\usepackage{latexsym}
\usepackage{amssymb}
\usepackage{longtable}
\usepackage{graphicx}
\usepackage{braket}
\usepackage{mathrsfs}

\usepackage{euscript,amssymb}
\usepackage{amsfonts}
\usepackage[dvips]{color}
\usepackage{color}

\def\del#1#2{\frac{\partial #1}{\partial #2}} 
\def\vec#1{\mathbf{#1}}
\def\be{\begin{equation}}
\def\ee{\end{equation}}
\def\D{\mathrm{d}}
\def\E{\mathrm{e}}
\def\I{\mathrm{i}}

\newcommand{\leftsuperindex}[2]{  { {}^{\mbox{\tiny{$(#2)$}}} \! #1 } }
\newcommand{\three}[1]{ \leftsuperindex{#1}{3} }
\newcommand{\four}[1]{ \leftsuperindex{#1}{4} }

\newcommand\hairspace{\kern .08333em }
\newcommand{\Real}{\Re\!\hairspace\mathfrak{e}\hairspace}
\newcommand{\Imag}{\Im\mathfrak{m}\hairspace}

\begin{document}

%%%%%%%%%%%%%%%%%%%%%%%%%%%%%%%%%%%%%%%%%%%%%%%%%%%%%%%%%%%%%%%%%%%%%%%%

\title{Quantum-gravitational effects on gauge-invariant scalar and tensor perturbations during inflation: The de Sitter case}

\author{David Brizuela}
\email{david.brizuela@ehu.eus}
\affiliation{Fisika Teorikoa eta Zientziaren Historia Saila, UPV/EHU,
  644 P.K., 48080 Bilbao, Spain} 

\author{Claus Kiefer}
\email{kiefer@thp.uni-koeln.de}
\affiliation{Institut f\"ur Theoretische Physik, Universit\"{a}t zu
  K\"{o}ln, Z\"{u}lpicher Stra\ss e 77, 50937 K\"{o}ln, Germany} 

\author{Manuel Kr\"amer}
\email{m.kraemer@wmf.univ.szczecin.pl}
\affiliation{Instytut Fizyki, Uniwersytet Szczeci\'{n}ski,
  ul.~Wielkopolska 15, 70-451 Szczecin, Poland} 

\date{\today}

%%%%%%%%%%%%%%%%%%%%%%%%%%%%%%%%%%%%%%%%%%%%%%%%%%%%%%%%%%%%%%%%%%%%%%%%

\begin{abstract}
We present detailed calculations for quantum-gravitational corrections  
to the power spectra of gauge-invariant scalar and tensor perturbations
during inflation. This is done by performing a semiclassical
Born--Oppenheimer type of approximation to the Wheeler--DeWitt
equation, from which we obtain a Schr\"odinger equation with
quantum-gravitational correction terms. As a first step, we perform our calculation for a de Sitter universe and find that the correction terms lead to an enhancement of power on the largest scales.
\end{abstract}

 \pacs{04.60.Bc, 04.60.Ds, 98.80.-k, 98.80.Qc}

\maketitle

\section{Introduction}

The unification of quantum theory and gravity is one of the central
open problems in physics. Several approaches to a theory of
quantum gravity have been developed \cite{oup}, but in order to
ultimately decide 
which approach describes nature best, we need testable
predictions. Finding such predictions is problematic, because
quantum-gravitational effects might only become sizable at energies of
the order of the Planck scale. A promising scenario to look for such
effects is the highly energetic inflationary phase in the very early
universe. This is also the topic of the present paper. 

In our work, we will focus on a rather conservative approach to
quantum gravity based on a canonical quantization that leads to the
Wheeler--DeWitt equation. We study how quantum fluctuations of
spacetime are influenced by quantum-gravitational effects during
inflation and whether the resulting corrections to the power spectra
of primordial fluctuations can be observed in the Cosmic Microwave
Background (CMB). Small effects may constitute the key to quantum
gravity, comparable perhaps with the discovery of the Lamb shift in
atomic physics \cite{KK12}. Canonical quantum gravity and the Wheeler--DeWitt equation
may not yield the ultimate quantum theory of
gravity, but they should give a reliable picture at least close to the
Planck scale; the Wheeler--DeWitt equation is the quantum wave
equation that directly gives the Einstein equations in the
semiclassical limit \cite{oup}. 

Quantum field theory in curved spacetime follows from the
Wheeler--DeWitt equation in a Born--Oppenheimer type of approximation
scheme \cite{oup}. This scheme may be realized by a formal expansion
with respect to the Planck mass \cite{singh,honnef}, but it can be realized
also in another way that is closer to the standard Born--Oppenheimer approximation
in molecular physics
\cite{bertoni}. Going one order beyond this limit yields the
quantum-gravitational corrections derived here. Previously, this
approximation was already used to derive such corrections from the
Planck-mass expansion
\cite{singh,honnef,BK98,KLM,KK-PRL,BEKKP,Gianluca,BE14} and from an 
alternative expansion \cite{ACT06,Sasha1,Sasha2,Sasha3}. 

In this paper, we continue with this investigation in the following ways. 
First, we apply the Planck-mass expansion scheme to the scalar and
tensor perturbations described in a fully gauge-invariant way.
Second, we keep all the terms in the resulting quantum-gravitationally
corrected Schr\"odinger equation for the perturbation modes and
explicitly address the issue of the appearing unitarity-violating terms.
We restrict ourselves here to the de Sitter case. This investigation
represents the preparation we need in order to expand our analysis to
the general case of slow-roll inflation, which we shall present in a
follow-up paper. 

The present paper is organized as follows. In section II, we first give an introduction to how
perturbations of spacetime are dealt with in a Hamiltonian framework. The quantization of this system
is discussed following the Dirac method. Then, the construction of perturbative gauge-invariant quantities
is presented, which leads to a trivialization of the perturbative constraints and allows for a reduced phase space
quantization. In section III, we then formulate the Wheeler--DeWitt equation
for our model of a perturbed inflationary universe. Sections II and III are presented in some detail because they contain
important reference material for later papers. Section IV is dedicated to the semiclassical approximation
of this Wheeler--DeWitt equation and it is at this point where we derive a Schr\"odinger equation with
quantum-gravitational correction terms for the quantized spacetime perturbations.
In section V we describe the Gaussian ansatz for the wave function representing
the perturbations and we use this in section VI to derive the general expression
for the power spectra of the scalar and tensor modes. We then present concrete results
for the usual (uncorrected) power spectra in a de Sitter universe in section VII. Finally
we calculate the quantum-gravitationally corrected power spectra for this case numerically
as well as -- after a linearization -- analytically in section VIII.
Section IX presents the main conclusions and comments on the comparison between the present
and previous approaches.

%%%%%%%%%%%%%%%%%%%%%%%%%%%%%%%%%%%%%%%%%%%%%%%%%%%%%%%%%%%%%%%%%%%%%%%%
\section{Hamiltonian perturbations in General Relativity}
\label{section:idea}
%%%%%%%%%%%%%%%%%%%%%%%%%%%%%%%%%%%%%%%%%%%%%%%%%%%%%%%%%%%%%%%%%%%%%%%%

\subsection{Classical framework}

Let us assume general relativity with the metric minimally coupled to a scalar matter field $\phi$.
The complete action of this system in the Hamiltonian form
can be written in the following way:
\begin{equation} \label{action}
{S} =
\int \D t \int \D^3x \left(\Pi^{ij} g_{ij,t} 
+ \pi_{\phi} \phi_{,t}-H\right),
\end{equation}
where $g_{ij}$ is the metric on the three-dimensional spatial slices defined
by the foliation given by the time function $t(x^\mu)$,
$\pi_{\phi}$ is the conjugate momentum of the scalar field $\phi$, and
$\Pi^{ij}$ is the conjugate momentum of the spatial metric. The Hamiltonian
density $H$ is a linear combination of constraints,
\begin{equation}
H:=N {\cal H} + N^i {\cal H}_i\,,
\end{equation}
where the lapse $N$ and the shift $N^i$ are Lagrange multipliers.
The metric of the four-dimensional $\four{g}_{\mu\nu}$
mani\-fold can be straightforwardly reconstructed in terms
of all these variables as follows,
\begin{equation}
N^{-2}= - \four{g}^{tt}\, , \qquad
N_i = \four{g}_{ti} \,, \qquad
g_{ij}=\four{g}_{ij} \,.
\end{equation}
In fact, these relations can also be understood as the definitions of
the lapse, shift, and the three-dimensional metric. 

The expressions for the Hamiltonian and diffeomorphism constraints are
respectively given by \cite{oup}, 
\begin{eqnarray}\label{hamiltonian}
{\cal H} &=& \frac{16\pi G}{\mu_g}
\left[\Pi^{ij}\Pi_{ij}-\frac{1}{2}\left(\Pi^l{}_l\right)^2\right]
     - \frac{\mu_g}{16\pi G} \three{R} \\
     &~&\;+\,\frac{1}{2} \left[\frac{\pi_{\phi}^2}{\mu_g}
                  +  \mu_g g^{ij} \phi_{,i}\phi_{,j}+2\, \mu_g {\cal V}(\phi)
                  \nonumber \right], \\
{\cal H}_i &=&  -\,2\kappa
\, D_j\Pi_i{}^j + \pi_{\phi} \phi_{,i}\,.
\end{eqnarray}
Here, and in the following, we have set $c=1$.
In these expressions, ${\cal V}(\phi)$ is the potential of the scalar field,
$\mu_g:=\sqrt{\det g_{ij}}$, $D$ is the covariant derivative
associated with $g_{ij}$, and all Latin indices are raised and lowered
by $g_{ij}$. The Einstein equations can be obtained by direct
variation of the action above. In particular, the constraints read
\begin{eqnarray}\label{eq_back1}
{\cal H}=0\,,\\
{\cal H}_i=0\,. \label{eq_back2}
\end{eqnarray}
The rest of the equations corresponds to the time derivatives of the metric,
the scalar field and their conjugate momentums. Nonetheless, their exact
form is not needed for our purposes, so they are not explicitly
displayed here.

Equations (\ref{eq_back1}) and (\ref{eq_back2}) are the general constraint
equations of canonical quantum gravity. In the following, however, we
will refer to them as the `background equations', and the total
constraint equations will consist of this background and the
perturbations. 

\subsection{Perturbations}\label{perturbations}

As a preparation for our application in the remaining part of the
paper, we will here and in the next subsection analyze the evolution of small
perturbations around a background 
that obeys the equations of the previous section. Let us assume that all quantities depend
on a dimensionless parameter $\varepsilon$, which defines a one-parameter family of spacetimes
$(M(\varepsilon), g_{\mu\nu}(\varepsilon),
T_{\mu\nu}(\varepsilon))$. The value $\varepsilon=0$ corresponds 
to the background spacetime.
In this way, a variational (or perturbative) operator $\delta$ can be defined as
\begin{equation}\label{def_delta}
\delta:=\left.\frac{\D}{\D\varepsilon}\right|_{\varepsilon=0}.
\end{equation}
Making use of this operator, the perturbations of different quantities
are named in the following way: 
\begin{eqnarray}\label{perturbative_variables}
C := \delta N \,, \quad &\qquad&
B^i := \delta(N^i) \,, \nonumber \\
h_{ij} := \delta(g_{ij}) \,, &\qquad& 
p^{ij} := \delta(\Pi^{ij}) \,, \nonumber \\
\varphi := \delta \phi \,, \quad &\qquad&
\quad p := \delta \pi_{\phi} \,.
\end{eqnarray}
The extremal value of the action, derived from
\begin{equation}
\delta S=0\,,
\end{equation}
provides the equations of motion for the background quantities.
Furthermore, as shown in \cite{Taub, Monc74}, the second variation of the action
gives an effective action functional for the perturbations,
\begin{eqnarray}\label{effaction}
\frac{1}{2}\,\delta^2{S} =
\int \D^4 x \biggl[ 
p^{ij} h_{ij,t} 
&+& p \,\varphi_{,t}
- C \,\delta({\cal H})
- B^i \,\delta({\cal H}_i) \nonumber \\
&-&\frac{N}{2} \,\delta^2({\cal H})
- \frac{N^i}{2} \,\delta^2({\cal H}_i)
\biggr] .
\end{eqnarray}
The first and second variations of the background constraints lead to
expressions $\delta({\cal H})$, $\delta({\cal H}_i)$, $\delta^2({\cal
  H})$, and $\delta^2({\cal H}_i)$ that are
explicitly given in e.g.~\cite{BrMa09}.

The equations for the small perturbations are then obtained by
variation of the action (\ref{effaction}) 
with respect to the perturbative variables
(\ref{perturbative_variables}). In particular, the variations with
respect 
to $B^i$ and $C$ give the constraints that must be obeyed by the perturbations:
\begin{eqnarray}\label{perturbative_constraint1}
\delta({\cal H})&=&0\,,\\\label{perturbative_constraint2}
\delta({\cal H}_i)&=&0\,.
\end{eqnarray}
The rest of the dynamical equations are obtained by taking the variations with
respect to $p^{ij}$, $h_{ij}$, $\varphi$, and $p$.

As mentioned in the previous section, in full general relativity the Hamiltonian
is just a linear combination of constraints and thus it vanishes on-shell (except for possible boundary terms). Nevertheless,
as shown here, for the small perturbations evolving on a given background
the situation is different. The Hamiltonian for the perturbations can be directly
read from (\ref{effaction}):
\begin{equation}
H_{\rm pert}:= C \,\delta({\cal H})
+ B^i \,\delta({\cal H}_i)
+\frac{N}{2} \,\delta^2({\cal H})
+ \frac{N^i}{2} \,\delta^2({\cal H}_i)\,.
\end{equation}
On-shell, due to the perturbative constraint equations (\ref{perturbative_constraint1}-\ref{perturbative_constraint2}), the
first two terms of the last equation vanish. But the last two terms of
this equation do not vanish and constitute a physical, non-zero, Hamiltonian
for the perturbative degrees of freedom:
\begin{equation}
H_{\rm phys}:=\frac{N}{2} \,\delta^2({\cal H})
+ \frac{N^i}{2} \,\delta^2({\cal H}_i)\,.
\end{equation}
Note that in this perturbative Hamiltonian the background quantities do not
appear as dynamical variables. That is, they are assumed to be given
and are not basic variables that encode
physical degrees of freedom. The system can be understood as a hierarchy of systems,
which can be solved order by order. The background quantities are obtained
as a solution to the background Einstein equations 
and are as such plugged in the action
(\ref{effaction}). Therefore, in this sense, 
the background quantities are non-dynamical in the Hamiltonian $H_{\rm
  pert}$.

\subsection{Quantization $\emph{\`a la Dirac}$}

In spite of the comments in the last paragraph, in order to quantize the system,
it is necessary to have a complete action for all degrees of freedom.
This can be constructed just by writing together the background action and
its second variation:
\begin{equation}
S_{\rm total}:= S + \frac{1}{2}\,\delta^2S\,,
\label{total_action}
\end{equation}
which defines the total Hamiltonian of the system:
\begin{eqnarray}\label{total_hamiltonian}
H_{\rm total}:= H &+& H_{\rm pert}=
N \left({\cal H}+\frac{1}{2} \,\delta^2{\cal H}\right)
\\&+& N^i \left({\cal H}_i + \frac{1}{2} \,\delta^2{\cal H}_i\right)
+C \,\delta{\cal H}
+ B^i \,\delta{\cal H}_i\,. \nonumber
\end{eqnarray}
Everything seems fine up to this point: the variation of the total action (\ref{total_action})
with respect to the perturbative variables (\ref{perturbative_variables}) gives the
linearized Einstein equations of motion.
This is because $S$ does not depend on the perturbative variables and, therefore, the
variation of $S_{\rm total}$ with respect to the perturbative variables is exactly the same
as the variation of $\frac{1}{2}\delta^2S$.
Nonetheless, $\frac{1}{2}\delta^2S$ does depend on background quantities. Thus,
if one takes the variations of the full action $S_{\rm total}$ with
respect to the background 
quantities in order to obtain their equations of motion, one will get
the background equations 
plus certain correction terms quadratic in the perturbations.
In particular, regarding the constraints, instead of obtaining
equations (\ref{eq_back1}--\ref{eq_back2}) 
we will get
\begin{eqnarray}
{\cal H}+\frac{1}{2}\,\delta^2{\cal H}=0\,,\\\label{diffconstraint2}
{\cal H}_i+\frac{1}{2}\,\delta^2{\cal H}_i=0\,.
\end{eqnarray}
Therefore, by using this full action to describe our physical system,
we are assuming that the background Einstein
equations are not exactly obeyed. Nonetheless, one could obtain the
exact background and perturbative equations from the total Hamiltonian
(\ref{total_hamiltonian}) by truncating the system at $\varepsilon$ order.

At a classical level, the discussion above is not very relevant since,
as already mentioned, 
all the properties and results we can derive from this framework (by
truncating the expansion 
of the action at a quadratic level) can only be trusted at $\varepsilon$ order.
Nonetheless, this has a great importance when quantizing this
system. We will assume 
a canonical quantization, where the wave function will depend on
position variables. 
All ``positions'' $(g_{ij},h_{ij},\phi,\varphi)$ and their corresponding conjugate momenta
$(\Pi_{ij},p_{ij},\pi_{\phi},p)$ will be promoted to operators, the first ones acting
multiplicatively and the latter ones as derivatives of their corresponding conjugate positions.

Therefore, the situation can now be interpreted in two alternative ways.
On the one hand, if one assumed that the equations of motion should be truncated at linear order,
and that the background equations are exactly obeyed, as we explained in the preceding section,
the constraints to be imposed on the wave function would be:
\begin{eqnarray}
\widehat{{\cal H}}\Psi&=&0\,,\\
\widehat{{\cal H}}_i\Psi&=&0\,,\label{diff_constraint}\\
\widehat{\delta{\cal H}}\Psi&=&0\,,\label{quantizeddeltaH}\\
\widehat{\delta{\cal H}}_i\Psi&=&0\,.
\end{eqnarray}
In this case there is, in addition, also a physical Hamiltonian
$H_{\rm phys}$ for some of
the degrees of freedom. This Hamiltonian would in principle lead to a Schr\"odinger
equation:
\begin{equation}
\label{Schroedinger}
\widehat{H}_{\rm phys}\Psi={\rm i}\,\frac{\partial\Psi}{\partial t}\,,
\end{equation}
where the time parameter $t$ should be defined from the background variables.
Nonetheless, this approach does not provide a direct interaction
between the quantized 
background and perturbations in the sense that each obeys its own
equations of motion. 

Note that if in equations (\ref{quantizeddeltaH}--\ref{Schroedinger})
only the perturbative degrees of freedom were quantized, whereas for
the background variables their 
classical behavior were considered, one would obtain the limit of quantum field
theories on a fixed background spacetime. In fact, a Schr\"odinger equation
like (\ref{Schroedinger}) results in a natural way from the Born--Oppenheimer
scheme presented below in that limit. But the important point is that
we want to go beyond that limit and for this purpose the dynamical
behavior of the background variables is important. The
quantum-gravitational corrections derived in our paper come from the
interaction of ``perturbations'' and ``background''. 

In order to 
obtain such a \emph{direct interaction} between the quantized background
and the perturbative degrees of freedom, 
one should directly quantize all degrees of
freedom with the Hamiltonian (\ref{total_hamiltonian}).
In this approach, there is no physical (non-vanishing) Hamiltonian, 
and one would have to impose the following constraints on the wave function:
\begin{eqnarray}
\label{ham_constraint}
\left(\widehat{{\cal H}}+\frac{1}{2}\,\widehat{\delta^2{\cal
      H}}\right)\Psi&=&0\,, \\\label{diff_constraint2}
\left(\widehat{{\cal H}}_i+\frac{1}{2}\,\widehat{\delta^2{\cal
      H}}_i\right)\Psi&=&0\,,\\ 
\widehat{\delta{\cal H}}\Psi&=&0\,, \label{quantum_constraint1}\\
\widehat{\delta{\cal H}}_i\Psi&=&0\,.\label{quantum_constraint2}
\label{deltaham_constraint}
\end{eqnarray}
This latter one is the method followed in \cite{HaHa85} and many other papers in order to
go beyond the approximation of quantum field theories on fixed backgrounds (see in particular the treatment in \cite{CK87}).
We will also work within this approach because it gives a consistent
formulation in the quantum theory.

The problem is that equations (\ref{ham_constraint}--\ref{deltaham_constraint})
are extremely difficult to be solved even for simple cosmological models. Therefore,
before quantizing the constraint equations, one can perform a convenient canonical
transformation to a new set of variables for which the linearized
constraints (\ref{quantum_constraint1}) and
(\ref{quantum_constraint2}) take a
much simpler form or, in the best of the cases, they are automatically obeyed.
This can be done by constructing so-called master gauge-invariant
variables, to which we now turn.

\subsection{Construction of master gauge-invariant quantities:
 reduced phase space quantization}\label{sec_gaugeinvariants} 

The constraints (\ref{perturbative_constraint1}--\ref{perturbative_constraint2})
are the generators of the perturbative gauge freedom. Instead of quantizing the whole
system $\emph{\`a la Dirac}$, as described in the previous section, one could solve
some of these perturbative constraints classically and perform a reduced phase
space quantization. One way to solve or, better to say, trivialize these constraints
is by constructing gauge-invariant quantities which encode the complete physical
information of the problem.

In order to explain how to construct gauge invariants in this Hamiltonian
setting, we will restrict the matter sector to the scalar field $\phi$ that we are considering.
Nonetheless, the generalization of the procedure to any other form of matter is straightforward.
Just for this section, let us denote by $(h_I,p_I)$, for $I=1,\dots,6$, the six independent components of the geometric
perturbative variables $(h_{ij}, p^{ij})$. Apart from those, we also have another perturbative degree
of freedom corresponding to the matter field, which is described by the pair $(\varphi,p)$.

The idea is essentially to make a canonical transformation from this
set of perturbative variables to another new set 
$(\tilde{h}_I,\tilde{p}_I,\tilde\varphi,\tilde p)$,
with the requirement that four of the new moments reproduce
 the constraints \eqref{perturbative_constraint1} and
 \eqref{perturbative_constraint2}: 
\begin{equation}\label{exampleconst}
\delta{\cal H}_J=\tilde{p}_J\,.
\end{equation}
The subindex $J$ takes the values $J=1,2,3,4$ and we have defined,
in order to have a compact notation for this section,
$\delta{\cal H}_4:=\delta{\cal H}$\,.
In this way, the four variables $\tilde{p}_J$ will be gauge-invariant, but constrained to vanish by equations (\ref{perturbative_constraint1}--\ref{perturbative_constraint2}),
whereas their conjugate variables $\tilde{h}_J$ will be pure gauge,
such that their initial value can be arbitrarily chosen.
In addition, the evolution equations for these latter variables
will contain the free functions $B^i$ and $C$, since they will be
obtained by variation of the action with respect to the linearized
constraints (\ref{exampleconst}). This allows us to choose also
the time derivatives of the pure-gauge variables $\tilde{h}_J$.

Therefore, if one were able to follow this procedure explicitly, one would have
isolated the non-physical information (gauge as well as constrained)
in the four pairs of conjugated variables $(\tilde{h}_J,\tilde{p}_J)$.
The remaining variables
$[(\tilde{h}_5,\tilde{p}_5),(\tilde{h}_6,\tilde{p}_6),
(\tilde\varphi,\tilde p)]$ 
will be the so-called {\em master variables}, which are gauge-invariant
and obey non-constrained equations of motion. These master variables
are the two degrees of freedom of the gravitational wave and
the matter degree of freedom, which contain all the physical information of the problem.
The initial variables $(h_I,p_I,\varphi, p)$ can be reconstructed in terms
of the master variables in any gauge just by applying the inverse
of the canonical transformation.

Examples of such master variables are the Regge--Wheeler \cite{ReWh57} and Zerilli \cite{Zeri70}
variables in Schwarzschild backgrounds, as well as the generalization of the former one
to dynamical spherical backgrounds \cite{GeSe79} (see \cite{Monc74, BrMa09, Bri15} for
explicit examples of the application of the described procedure for spherical backgrounds).
In cosmological backgrounds, the decomposition into scalar, vector, and
tensor quantities is usually performed \cite{Bar80, KoSa84}. In this decomposition, the gravitational
degrees of freedom are encoded into the tensorial sector and they are
automatically gauge-invariant because they do not appear in the perturbative
constraints (\ref{perturbative_constraint1}--\ref{perturbative_constraint2}). (Note that the constraints form a
four-dimensional vector field and its decomposition gives rise to two scalar
and two vector degrees of freedom, but no tensor components.)
For a matter sector given by a scalar field, the gauge-invariant
Mukhanov--Sasaki variable \cite{Muk88,Mukhanov:1992}, see below, can be defined by following the
described procedure \cite{Gun93}. A different technique, inspired by
Hamilton--Jacobi theory, which leads to the same result for such a cosmological
background, can be found in \cite{Langlois}.

Once the constraints at the linearized level are trivialized, one can
either solve them 
classically by imposing the four moments $\tilde p_J$ to be vanishing
or quantize them. The quantization would just tell us, through equations (\ref{quantum_constraint1}--\ref{quantum_constraint2}), that the wave function must be independent of the position variables $\tilde h_J$ \cite{Gun93}.
Therefore, only equations (\ref{ham_constraint}--\ref{diff_constraint2}) will have to be considered.
Furthermore, in spatially homogeneous backgrounds, as the one that will be considered in this paper,
by choosing adapted coordinates with a vanishing shift function $N_i$, the background diffeomorphism constraint
(\ref{eq_back2}) is exactly obeyed and equation (\ref{diffconstraint2}), and thus its quantum version
(\ref{diff_constraint2}), do not appear in this case [see (\ref{total_hamiltonian})].
Thus, by constructing gauge-invariant variables at the classical level following the commented procedure,
only the Hamiltonian constraint (\ref{ham_constraint}) will have to be solved at the quantum level.
The total gauge-invariant Hamiltonian will take the following form:
\begin{eqnarray}\label{total_GI_ham}
\int \D^3x\,\, H_{\rm total}^{(\text{gi})}&=&\int \D^3x \,N\left({\cal H}+\frac{1}{2}\,\delta^2{\cal H}\right)\nonumber\\
&=&{}^0{\cal H} +{}^\text{S}{\cal H}+{}^\text{T}{\cal H}\,, 
\end{eqnarray}
where ${}^0{\cal H}$ will be the (spatially integrated) Hamiltonian corresponding to
the background spacetime, see (\ref{backgroundH}) below,
 while ${}^\text{S}{\cal H}$ and ${}^\text{T}{\cal H}$
will be the Hamiltonians of the scalar, see (\ref{eq:hamilton}), and
tensorial, see (\ref{H_tensors}),
perturbative degrees of freedom, respectively. The quantization
of this constraint will lead to (\ref{ham_constraint}), which, for our specific model,
will take the explicit form given in (\ref{eq:WdWmaster}). This equation will be called
the master Wheeler--DeWitt equation and the main objective of this paper will be
to solve it within a semiclassical Born--Oppenheimer approach;
therein, \eqref{Schroedinger} will emerge as an approximate equation.

Here, two comments are in order. On the one hand, note that the procedure described in this
section, which was also similarly employed in \cite{Sasha1}, only takes into account the perturbative degrees of freedom. We are assuming
that the time derivatives of the background variables that appear during the process
should be replaced by making use of the background equations of motion. This is sufficient
for our approach since, as will be explained below, for the background objects that appear
inside ${}^\text{S}{\cal H}$ and ${}^\text{T}{\cal H}$, their classical behavior will be considered.
If one wishes to perform a more fundamental quantization, without making use of the
background equations of motion, it would be necessary
to complete the above-mentioned canonical transformation to the background degrees of
freedom by correcting them with quadratic terms in the perturbations along the lines presented
in \cite{GMM15, PP07}. Nonetheless, note that both procedures would lead to the same Hamiltonian \eqref{total_GI_ham}.

On the other hand, the construction of the master
gauge-invariant objects, and their corresponding Hamiltonian,
is quite involved and has already been performed in \cite{Gun93, Langlois}
for the background spacetime that we will be interested in. Since we already know which are the
canonical variables, it is thus possible to use a reduced
action approach \cite{DGP92}, as originally performed in \cite{Muk88},
where the constraints are solved and used to simplify the action,
to obtain the Hamiltonian (\ref{total_GI_ham}). This will be
explicitly presented in the next 
section in order to clarify our notation and assumptions.

\section{The Wheeler--DeWitt equation}

In this section, the formalism explained in the previous section
will be applied to the study of an inflationary universe with tiny
perturbations that are interpreted as the seeds for structure
formation in the universe and, hence, are in particular responsible
for the anisotropies of the CMB. 
In our model, the inflationary phase is caused by a scalar
\emph{inflaton} field that only varies slightly with time. We consider
scalar and tensor perturbations of the metric, as well as perturbations
of the scalar field. The resulting Wheeler--DeWitt equation is then
the starting point to calculate the power spectra of scalar and tensor
perturbations by using a semiclassical approximation.

\subsection{The minisuperspace background}

We start by formulating the Wheeler--DeWitt equation for a flat
Friedmann--Lema\^itre--Robertson--Walker (FLRW) universe
with spatial topology. This plays the
role of our background spacetime, on which the perturbations are formulated.
But we emphasize that, as described in the previous section and contrary to the
usual approach of quantum field theory on fixed backgrounds, this background
will also be treated dynamically with quantum degrees of freedom. 

 We restrict ourselves to a flat universe,
because the exponential expansion during the inflationary phase
flattens any curvature of the universe sufficiently fast. In terms of
the conformal time $\eta$, which is fixed by choosing the lapse
function as $N=a$, $a$ being the scale factor, the line
element of such a universe reads 
\be
\D s^2 = a^2(\eta)\left(-\,\D\eta^2 + \D\vec{x}^2\right).
\ee
As usual, the relation between this conformal time and the cosmic time $t$ is given by
${\D\eta}/{\D t} = a^{-1}$.
As mentioned above, we introduce a massive minimally coupled
scalar field $\phi$ with
potential $\mathcal{V}(\phi)$, which plays the role of the inflaton, and
we neglect a cosmological constant. The action for such a
model takes the form 
\be \label{SinclL}
S = \frac{1}{2}\int
\D\eta\,\mathfrak{L}^3\left[-\,\frac{3}{4\pi G}\,(a')^2 +
  a^2\,(\phi')^2-2\,a^4\,\mathcal{V}(\phi)\right].
 \ee
Here and in the following, derivatives with respect to $\eta$ are denoted using
primes. For completeness, let us include explicitly the equations of motion, which
are easily obtained from this action:
\begin{align}
\frac{3}{4\pi G}\,\frac{a''}{a}+(\phi')^2-4\,a^2\,{\cal V}(\phi)&=0\,,\label{friedmanneq}\\
\phi''+2\,\frac{a'}{a}\,\phi'+a^2\,\frac{\D {\cal V}(\phi)}{\D\phi}&=0\,.
\end{align}
In \eqref{SinclL}, we have introduced an arbitrary
length scale $\mathfrak{L}$ that appears when performing the integration
over the volume. In order not to explicitly
denote this length scale in the following, we rescale the scale factor
and conformal time as follows, which is similar to the replacement
carried out in \cite{Sasha2}, 
\be \label{repl1}
a_\text{new} = a_\text{old}\,\mathfrak{L}\,, \qquad \eta_\text{new} =
\frac{\eta_\text{old}}{\mathfrak{L}}\,. 
\ee
After this replacement, the scale factor $a$ has the dimension
of a length, while $\eta$ -- as well as the spatial variables --
become dimensionless. Hence, we can
effectively set the length scale $\mathfrak{L}$ appearing in
\eqref{SinclL} to one. However, later in our discussions of possible
observational consequences of our results, we will restore
$\mathfrak{L}$.
With this replacement, the Hamiltonian for our `background universe'
 reads as follows:
\be\label{backgroundH}
{}^0{\cal H}
= -\,\frac{2\pi G}{3}\,\pi_a^2 + \frac{1}{2a^2}\,\pi_\phi^2
+ a^4\,\mathcal{V}(\phi)\,,
\ee
where $\pi_a=-3a'/(4\pi G)$ and $\pi_{\phi}=a^2\phi'$.
The canonical quantization of this Hamiltonian is carried out using
the Laplace--Beltrami factor ordering \cite{oup}, which 
guarantees that the kinetic term remains invariant under transformations in
configuration space. Consequently, we obtain the following 
Wheeler--DeWitt equation for the background: 
\be
\Biggl[\frac{2\pi\hbar^2 G}{3 a}\,\del{}{a}\left(a\,\del{}{a}\right) -
\frac{\hbar^2}{2a^2}\,\del{^2}{\phi^2}+a^4\,\mathcal{V}(\phi)\Biggr]\Psi_0(a,\phi)
= 0\,. 
\label{WdW_raw_C5}
\ee
We simplify this equation by introducing the dimensionless quantity $\alpha$, which
is defined in terms of a reference scale factor $a_0$ as 
\be
\label{alpha}
\alpha := \ln\!\left(\frac{a}{a_0}\right).
\ee
Furthermore, we set $\hbar = 1$ and define for convenience a rescaled Planck mass
\be \label{mpdef}
m_\text{P}^2 := \frac{3}{4\pi G}\,.
\ee
With these definitions, the Wheeler--DeWitt
equation simplifies considerably and reads 
\be
\label{WdW}
\frac{1}{2}\,\E^{-2\alpha}\Biggl[\frac{1}{m_\text{P}^2}\,\del{^2}{\alpha^2}-
\del{^2}{\phi^2}+2a_0^6\E^{6\alpha}\,\mathcal{V}(\phi)\Biggr]\Psi_0(\alpha,\phi) 
= 0\,. 
\ee

\subsection{Scalar perturbations} \label{sectpert}

We now introduce scalar perturbations to the background metric and
parametrize them using four scalar functions of space and time, $A$,
$B$, $\psi$ and $E$. The perturbed metric then takes the form
\begin{eqnarray}
\label{eq:metric}
\D s^2=a^2(\eta)\Bigl\lbrace
&-&\left(1-2A\right)\D\eta^2
+2\left(\partial_iB\right)\D x^i\D \eta \nonumber\\
&+&\left[\left(1-2\psi\right)\delta_{ij}
+2\partial_i\partial_jE
\right]\D x^i\D x^j\Bigr\rbrace\,.\quad
\end{eqnarray}
Here, we still take $a$ to be dimensionless and $\eta$ to have the
dimension of a length; the redefinitions using $\mathfrak{L}$ will be
done below. 

As has been explained in section \ref{sec_gaugeinvariants},
the perturbation of the scalar field $\varphi:=\delta\phi$, together with the four scalars
in the metric (\ref{eq:metric}), can be combined into {\em one}
gauge-invariant master scalar which is sufficient to fully describe the scalar
perturbations \cite{Mukhanov:1992}.
 We will construct such a gauge-invariant
quantity using one of the so-called \emph{Bardeen potentials},
$\Phi_{\rm B}$, which is defined by
\be
\label{eq:defbardeen}
\Phi_{\rm B}(\eta,\vec{x}) :=
A+\frac{1}{a}
\left[a\left(B-E^\prime\right)\right]^\prime.
\ee
Note that $\Phi_{\rm B}$ is dimensionless. 
On top of the metric perturbations, we also need to consider
fluctuations $\varphi(\eta,\vec{x})$ of the scalar inflaton field
$\phi(\eta)$, which can be represented in a gauge-invariant way as
\be
\varphi^{(\mathrm{gi})}(\eta,\vec{x}):=
\varphi+\phi^\prime\left(B-E^\prime\right)\, .
\ee
The quantities $\Phi_{\rm B}$ and $\varphi^{(\mathrm{gi})}$ can be
combined to a single master gauge-invariant quantity, the so-called
\emph{Mukhanov--Sasaki variable} $v$, which gives a full description
of the scalar sector of the perturbations \cite{Mukhanov:1992}. This
variable is defined as
\be \label{Mukhdef}
v(\eta,\vec{x}):=a
\left[\varphi^{(\mathrm{gi})}
+\phi^\prime\frac{\Phi_{\rm B}}{\mathscr{H}}\right],
\ee
where $\mathscr{H} := a'/a=Ha$, and $H=\dot{a}/a$ is the standard Hubble parameter. 
Here, $v$ has the dimension of an inverse length. Including the scale factor $a$ in
\eqref{Mukhdef} is essential, because only that way $v$ becomes the canonical variable
that diagonalizes the action, that is, there are no cross terms of the
form $v' v$, and therefore 
no first-order derivative terms appear in its equation of motion. In this way, the field $v$
obeys a Klein--Gordon equation on a Minkowski background. In quantum field theory on
classical backgrounds, this is the way one always writes the action, such that one can
impose the usual definition of the creation and annihilation operators, with their canonical
commutation relations and the usual Klein--Gordon inner product \cite{DGP92}.

Apart from this, there are two other arguments in favor of introducing
the variable \eqref{Mukhdef}. First,
the employed re-scaling with $a$ in (\ref{Mukhdef}) is also
crucial in describing the quantum-to-classical transition
by the process of decoherence: it has been shown that 
the divergence occurring in the decoherence factor can successfully be
regularized when using this re-scaling \cite{BKKM99}. 
Second, this re-scaling allows for a unitary implementation of the
quantum dynamics when the perturbations are treated for quantum field
theory in curved spacetime in the Fock picture \cite{Fock}.

In order to obtain the action for the variable $v$, one has to expand the Einstein--Hilbert action plus the scalar field action up to the second order in the perturbations around a FLRW background, and one ends up with the
scalar part of the second variation of the action \cite{Mukhanov:1992} 
\be
\label{eq:action}
\frac{1}{2}\,\delta^2 S=\frac{1}{2}
\int{\D\eta\,\D^3 \vec{x}
\left[\left(v^\prime\right)^2
-\delta^{ij}\,\partial_iv\,\partial_jv
+\frac{z^{\prime\prime}}{z}\,v^2\right]}\, .
\ee
Here, $z$ is a shorthand notation for
\be\label{defz}
z:=a\sqrt{\epsilon}\,,
\ee
where we have used the first slow-roll parameter
\be \label{defepsilon}
\epsilon := -\,\frac{\dot{H}}{H^2} = 1- \frac{\mathscr{H}'}{\mathscr{H}^2}\,.
\ee
The fraction $z''/z$ is therefore in the general case a complicated expression that contains up to fourth-order $\eta$-derivatives of
the background variables $\phi$ and $a$. 
Given that we are dealing here with a full theory of quantum gravity, where the background is quantized as well, we would therefore in principle also have to replace the $\eta$-derivatives appearing in this fraction, which will become canonical momenta in the respective Hamiltonian, by derivatives with respect to $a$ and $\phi$ in the process of quantization.
It is, however, a feature of the semiclassical (Born--Oppenheimer) approximation in quantum cosmology that the
minisuperspace momenta can be approximated, at the used order, by derivatives of the minisuperspace variables \cite{CK87}. 
Therefore, we will use the classical expressions for $z''/z$, which leads to a drastic simplification when considering the de Sitter and slow-roll cases.

Since we are working in the realm of linear perturbation theory, we assume that each mode of the perturbations evolves independently, which allows us to perform a Fourier transform of the variable $v$ as follows:
\be
\label{eq:tfv}
v\left(\eta,\vec{x}\right)= \int_{\mathbb{R}^3}\frac{\D^3 \vec{k}}{\left(2\pi\right)^{3/2}}\,
v_{\vec{k}}(\eta)\,
\E^{\I\vec{k}\cdot \vec{x}}\, .
\ee
The wave vector $\vec{k}$ that appears in this equation and its modulus $k = |\vec{k}|$ are related to a respective wave length \emph{without} including the factor $2\pi$, such that the relation between $k$ and its wave length $L$ is simply given by
\be
k = L^{-1}\,.
\ee
We assume that $v$ is real, such that the relation $v_{-\vec{k}}=v_{\vec{k}}^*$ holds. The action \eqref{eq:action} takes the following form after applying the Fourier transform and taking the integral over the spatial volume:
\be
\label{eq:actionfourier}
\frac{1}{2}\,\delta^2 S= \frac{1}{2}\int\D \eta\int \D^3 \vec{k}
\left\lbrace v_{\vec{k}}^\prime{v_{\vec{k}}^*}^\prime
+v_{\vec{k}}v_{\vec{k}}^*\left[
\frac{z^{\prime\prime}}{z}
-k^ 2\right]\right\rbrace .
\ee
In order to later be able to analyze each mode separately, that is, to
have a separate Wheeler--DeWitt equation for each mode, we have to
replace the integral over the wave vector $\vec{k}$ by a sum. 
This can be justified if one uses, as we do here, a compact spatial
topology. In order to implement this formally, we introduce an
arbitrary length scale 
$\mathfrak{L}$, with respect to which the wave modes are discretized,
in the following way (see e.g. \cite{Parker:2009}, Appendix) 
\be \label{inttodiscr}
\int \D^3\vec{k} \, \biggl\{\,\cdots\biggr\}\quad\rightarrow\quad
\frac{1}{\mathfrak{L}^3}\sum_{\vec{k}}\, \biggl\{\,\cdots\biggr\}\,. 
\ee
Here a comment is in order. Both the equations of motion for the
perturbations and the initial data that we will later consider will
only depend on the module $k$. Therefore, the dependence 
of all results on the vector ${\vec k}$ will be only through its
module and, in particular, the sum in \eqref{inttodiscr} has to be
taken over the module $k$. Nevertheless, in order to keep the notation
as much general as possible we will leave the notation for all
quantities as ${\vec k}$. 

Our action therefore becomes
\be
\label{eq:actionfourier2}
\frac{1}{2}\,\delta^2 S= \frac{1}{2}\int\D \eta\,\frac{1}{\mathfrak{L}^3}\sum_{\vec{k}}
\left\lbrace v_{\vec{k}}^\prime{v_{\vec{k}}^*}^\prime
+v_{\vec{k}}v_{\vec{k}}^*\left[
\frac{z^{\prime\prime}}{z}
-k^ 2\right]\right\rbrace .
\ee
Like in the previous section, we want to eliminate $\mathfrak{L}$ and only reintroduce it when comparing our results with observations. Following a procedure similar to \cite{Sasha2}, we thus apply the replacements given in \eqref{repl1} and additionally replace $v$ and $k$ as follows:
\be \label{repl2}
\quad v_\text{new} = \frac{v_\text{old}}{\mathfrak{L}^2}\,, \quad k_\text{new} = k_\text{old}\,\mathfrak{L}  \,.
\ee
The replacements imply that the wave vector $k$ is now regarded as a dimensionless quantity until we introduce a reference scale later in the observational quantities. 
The action thus reads
\be
\label{eq:actionfourier3}
\frac{1}{2}\,\delta^2 S
= \frac{1}{2}\int\D \eta\,\sum_{\vec{k}}
\left\lbrace v_{\vec{k}}^\prime{v_{\vec{k}}^*}^\prime
+v_{\vec{k}}v_{\vec{k}}^*\left[
\frac{z^{\prime\prime}}{z}
-k^ 2\right]\right\rbrace .
\ee
Note that all the variables that appear here are dimensionless (in the system
of units with $\hbar=1=c$ that we use here).  
As usual, from this Lagrangian we now define the canonical momenta as
\be
\pi_{\vec{k}}=v_{\vec{k}}^\prime
\ee
and therefore end up with the Hamiltonian
\be
\label{eq:hamilton}
{}^\text{S}{\cal H}
=\frac{1}{2}\,\sum_{\vec{k}}
\left\lbrace \pi_{\vec{k}}\pi_{\vec{k}}^*
+v_{\vec{k}}v_{\vec{k}}^*
\left[k^2-
\frac{z^{\prime\prime}}
{z}\right]\right\rbrace,
\ee
which is also a dimensionless quantity. 
As already mentioned above, we should also replace the $\eta$-derivatives of $a$ and $\phi$ that appear in $z''/z$ by their canonical momenta, in order to perform a complete quantization including the background variables $a$ and $\phi$ as, for example, shown in \cite{Langlois}. 
However, since this would lead to a significantly more complicated quantization procedure and given that we will later only consider the de Sitter and, in a subsequent paper, the slow-roll case, where $z''/z$ is approximated by a small value of the slow-roll parameter $\epsilon$, determined at the
classical level, we will treat $z''/z$ as a classical quantity. Also note that $z''/z$ would not change if we completed the canonical transformation to make the perturbation variables gauge-invariant to the background, as commented at the end of section \ref{sec_gaugeinvariants}.

To simplify the notation, we define the following quantity that one can regard as the 
time-dependent (dimensionless) frequency of the parametric harmonic oscillator described by the Hamiltonian \eqref{eq:hamilton}:
\be
\label{eq:defomega}
{}^\text{S}\omega^2_{\vec{k}}(\eta)=k^2-
\frac{z^{\prime\prime}}
{z}\,.
\ee
Before proceeding to the quantization, we should, in principle, define a new set of \emph{real} variables from $v_{\vec k}$ and $\pi_{\vec k}$ by creating a double copy of the variables as well as the wave function, as it is e.g.~presented in \cite{Martin:2012}. Without such a procedure, the quantization is not entirely consistent, as we will quantize terms like $v_{\vec{k}}v_{\vec{k}}^*$ as $v^2$. However, since it will not make any difference in the calculation later on, whether we employ the strict quantization procedure or not, we shall refrain from introducing such variables for the sake of briefness and clarity.

The quantization is thus carried out by promoting $v_{\vec k}$ and $\pi_{\vec k}$ to quantum operators $\hat{v}_{\vec k}$ and $\hat{\pi}_{\vec k}$ and demanding that they obey the standard commutation relations
\be
\left[\hat{v}_{\vec k},\hat{\pi}_{\vec p}\right]
=\I\,\delta\left({\vec k}-{\vec p}\right).
\ee
The operators $\hat{v}_{\vec k}$ and $\hat{\pi}_{\vec k}$ are then represented by the following expressions:
\be
\label{eq:ElemActions1}
\hat{v}_{\vec{k}}\Psi=v_{\vec{k}}\Psi\, ,\quad\hat{\pi}_{\vec{k}}\Psi=-\,\I\,\frac{\partial\Psi}{\partial v_{\vec{k}}}
\ee
and thus we finally obtain the quantum Hamiltonian for the scalar perturbations:
\be
\label{HamS}
{}^\text{S}{\widehat{\mathcal{H}}}=\sum_{\vec{k}} {}^\text{S}{\widehat{\mathcal{H}}_{\vec{k}}} = \sum_{\vec{k}}\left\{-\,\frac{1}{2}\,\frac{\partial^2}{\partial v_{\vec k}^2}+\frac{1}{2}\,{}^\text{S}\omega^2_{\vec k}(\eta)\,v_{\vec{k}}^2\right\}.
\ee

\subsection{Tensor perturbations}

We now want to introduce as well tensor perturbations of the metric. While scalar perturbations lead to temperature anisotropies in the CMB, tensor perturbations represent primordial gravitational waves and their presence would additionally lead to a polarization of the CMB radiation, which would be especially detected in the so-called B-modes (see e.g.~\cite{Seljak}).
We will focus here, however, on the influence the primordial gravitational waves have on the anisotropy spectrum. 

Tensor perturbations can be represented by a symmetric, traceless and divergenceless tensor $h_{ij}$ in the subsequent way:
\be
\D s^2=a^2(\eta)\left[-\,\D\eta^2+\left(\delta_{ij}+h_{ij}\right)\D x^i\D x^j\right].
\ee
By construction, tensor perturbations are already gauge-invariant. The symmetric tensor $h_{ij} = h_{ji}$ has six independent components, but due to its divergencelessness $\partial^i h_{ij}=0$, these are reduced by three and additionally by one due to its tracelessness $\delta^{ij}h_{ij}=0$. Hence, only two degrees of freedom remain and they correspond to the two polarizations of gravitational waves, $h^{(+)}$ and $h^{(\times)}$.

We define the perturbation variable for the tensor sector as
\be
v_{\vec{k}}^{(\lambda)} := \frac{a\,h^{(\lambda)}_{\vec{k}}}{\sqrt{16\pi G}}\,,
\ee
where $\lambda$ stands for the two polarizations $+$ and $\times$. Like for the scalar perturbations, $v_{\vec{k}}^{(\lambda)}$ thus includes the scale factor $a$ as prefactor. 
The derivation of the perturbation Hamiltonian is completely analogous to the scalar case, the only differences lie in the two polarizations present and in the form of the time-dependent frequency, which in the tensor case is given by the simpler expression
\be
\label{eq:defTomega}
{}^\text{T}\omega^2_{\vec{k}}(\eta)=k^2-
\frac{a^{\prime\prime}}
{a}\, .
\ee
Replacing the integral of the total Hamiltonian for the tensor perturbations by a sum using the method presented in \eqref{inttodiscr} and \eqref{repl2}, we end up with the subsequent Hamiltonian for the tensor perturbations,
\be\label{H_tensors}
{}^\text{T}{\mathcal{H}}=\sum_{\lambda=+,\times}
\sum_{\vec{k}}\left\{\frac{1}{2}\,\pi_{\vec k}^{(\lambda)} \pi_{-\vec k}^{(\lambda)}+\frac{1}{2}\,{}^\text{T}\omega^2_{\vec k}(\eta)\,v_{\vec{k}}^{(\lambda)} v_{\vec{-k}}^{(\lambda)}\right\}.
\ee
We see that this Hamiltonian is of essentially the same form as the Hamiltonian for the scalar perturbations, such that we can treat both cases together in one step. In particular, its quantization is directly given by
\begin{equation*}
{}^\text{T}{\widehat{\mathcal{H}}}=\sum_{\lambda}\sum_{\vec{k}} {}^\text{T}{\widehat{\mathcal{H}}_{\vec{k}}} = \sum_{\lambda;\vec{k}}\left\{-\,\frac{1}{2}\,\frac{\partial^2}{\partial v_{\vec k}^{^{(\lambda)}2}}+\frac{1}{2}\,{}^\text{T}\omega^2_{\vec k}(\eta)\,v_{\vec{k}}^{^{(\lambda)}2}\right\}.
\end{equation*}

\subsection{Master Wheeler--DeWitt equation}

Adding the quantum Hamiltonians for the scalar and tensor perturbations to the Wheeler--DeWitt equation of the background, equation (\ref{ham_constraint}) takes the form
\begin{widetext}
\be
\label{eq:WdWmaster1}
\frac{1}{2}\Biggl\{\E^{-2\alpha}\left[\frac{1}{m_\text{P}^2}\,\del{^2}{\alpha^2}-\del{^2}{\phi^2}+2\,\E^{6\alpha}\,\mathcal{V}(\phi)\right] \\
+ \sum_{\vec{k}}\left[-\,\frac{\partial^2}{\partial v_{\vec k}^2}+{}^\text{S}\omega^2_{\vec k}(\eta)\,v_{\vec{k}}^2\right] 
+ \sum_{\lambda;\vec{k}}\left[-\,\frac{\partial^2}{\partial v_{\vec k}^{^{(\lambda)}2}}+{}^\text{T}\omega^2_{\vec k}(\eta)\,v_{\vec{k}}^{^{(\lambda)}2}\right]
\Biggr\}\Psi(\alpha,\phi,\{v_{\vec{k}}\}) = 0\,.
\ee
\end{widetext}
Here and in the following, it is implicitly understood that the reference scale 
$a_0$ introduced in (\ref{alpha}) is associated with every factor of $\E^{\alpha}$.
In order to simplify the notation, we will, for now on, no longer write out the superscripts S and T, and we will also refrain from putting a circumflex over quantized quantities. Hence, in this notation, the last equation \eqref{eq:WdWmaster1} reads as follows,
\begin{align}
\label{eq:WdWmaster}
\frac{1}{2}\Biggl\{&\E^{-2\alpha}\left[\frac{1}{m_\text{P}^2}\,\del{^2}{\alpha^2}-\del{^2}{\phi^2}+2\,\E^{6\alpha}\,\mathcal{V}(\phi)\right] \\
&+ \sum_{\vec{k};\text{S},\text{T}_\lambda}\left[-\,\frac{\partial^2}{\partial v_{\vec k}^2}+\omega^2_{\vec k}(\eta)\,v_{\vec{k}}^2\right]\Biggr\}\Psi(\alpha,\phi,\{v_{\vec{k}}\}) = 0\,. \nonumber
\end{align}
Since we assume throughout this work that the perturbations are small and that the perturbation modes do not interact with each other, we can 
make the following product ansatz \cite{KK-PRL},
\be
\Psi\big(\alpha,\phi,\{v_{\vec{k}}\}\big) =
\Psi_0(\alpha,\phi)\prod_{\vec{k};\text{S},\text{T}_\lambda}\widetilde{\Psi}_{\vec{k}}(\alpha,\phi,v_{\vec{k}}).
\ee
We can then define a wave function for each mode $\vec{k}$ for \emph{both} scalar and tensor perturbations in the following way
\be
\Psi_{\vec{k}}(\alpha,\phi,v_{\vec{k}}):=
\Psi_0(\alpha,\phi)\widetilde{\Psi}_{\vec{k}}(\alpha,\phi,v_{\vec{k}})\,.
\ee
Each of these wave functions then obeys an individual Wheeler--DeWitt equation for its respective mode $\vec{k}$
\begin{align}
\label{eq:WdWk}
\frac{1}{2}\Biggl\{\E^{-2\alpha}\biggl[&\frac{1}{m_\text{P}^2}\,\del{^2}{\alpha^2}-\del{^2}{\phi^2}+2\,\E^{6\alpha}\,\mathcal{V}(\phi)\biggr] \nonumber \\
&-\,\frac{\partial^2}{\partial v_{\vec k}^2}+\omega^2_{\vec k}(\eta)\,v_{\vec{k}}^2\Biggr\}\Psi_{\vec{k}}(\alpha,\phi,v_{\vec{k}}) = 0\,,
\end{align}
with each corresponding frequency ${}^\text{S}\omega^2_{\vec k}(\eta)$, as given in \eqref{eq:defomega}, or ${}^\text{T}\omega^2_{\vec k}(\eta)$,
as defined in \eqref{eq:defTomega}.

In order to streamline the notation, we shall now introduce a minisuperspace metric.
For this purpose and with regard to the semiclassical approximation carried out later -- for which we want to have a common prefactor of $m_\mathrm{P}^{-2}$ in front of the derivative terms of the background variables -- we redefine the scalar field to make it dimensionless,
\be
\tilde{\phi} := m_\mathrm{P}^{-1} \phi\,.
\ee
We can thus define the minisuperspace variable $q^A$, 
which takes an index of either the value $0$ or $1$, in the following way
\be
q^0  := \alpha \quad \text{and} \quad q^1 := \tilde{\phi}\,.
\ee
The corresponding minisuperspace metric $\mathcal{G}_{AB}$ is then given by
\be \label{mssmet}
\mathcal{G}_{AB} := \text{diag}\!\left(-\,\E^{-2\alpha}, \E^{-2\alpha}\right),
\ee
and we introduce an auxiliary potential $V$, which has the dimension of length squared, via
\be \label{auxV}
V(q^A) := \frac{2}{m_{\rm P}^2}\,\E^{4\alpha}\,\mathcal{V}(\phi)\,.
\ee
The Planck mass squared that has been introduced in this definition
will be relevant in the next section
in order to construct a hierarchy of equations. Note that such a term can always be
introduced by rescaling the corresponding coupling constants of the potential. In the
particular case of quadratic potentials, $V$ could be written in terms of $\tilde{\phi}$ as
\be
V(q^A) = 2\,\E^{4\alpha}\,\mathcal{V}(\tilde{\phi})\,.
\ee
Using all these definitions, equation \eqref{eq:WdWk} finally reads
\begin{align}
\label{eq:WdWks}
\frac{1}{2}\Biggl\{-\,\frac{1}{m_\mathrm{P}^2}\,\mathcal{G}_{AB}\,&\del{^2}{q_A\partial q_B}+m_\text{P}^2\,V(q^A) \nonumber \\
-\,&\frac{\partial^2}{\partial v_{\vec k}^2}+\omega^2_{\vec k}(\eta)\,v_{\vec{k}}^2\Biggr\}\Psi_{\vec{k}}(\alpha,\phi,v_{\vec{k}}) = 0\,.
\end{align}
This equation will be referred to as our {\em master} Wheeler--DeWitt equation and it will be the starting point for the semiclassical approximation we carry out in the next section.

Note that strictly speaking, \eqref{eq:WdWks} is not a Wheeler--DeWitt equation in the original sense, because the conformal time $\eta$ appears explicitly in $\omega^2_{\vec k}(\eta)$ due to our chosen approximation. In principle, as mentioned before, one would have to include the minisuperspace momenta in $\omega^2_{\vec k}(\eta)$ instead. However, in order not to complicate the terminology, we will keep referring to \eqref{eq:WdWks} as our master Wheeler--DeWitt equation.

\section{Semiclassical approximation}
\label{sectsemi}

Instead of solving the master Wheeler--DeWitt equation \eqref{eq:WdWks} directly, we are going to apply the semiclassical approximation scheme that was first presented in \cite{singh} and applied to cosmology in \cite{KK-PRL}. The advantage of this approximation is that we recover, at consecutive orders,
first the dynamics of the classical background, then a Schr\"odinger equation for the perturbations propagating on the classical background and, finally,
quantum-gravitational corrections to it. In this way, we are able to clearly distinguish quantum-gravitational effects.

For our approximation, we use the following WKB-type ansatz
\be \label{WKBansatz}
\Psi_{\vec{k}}(q^A,v_{\vec{k}}) = \text{e}^{\text{i}\,S(q^A,v_{\vec{k}})}\,,
\ee
where the function $S(q^A,v_{\vec{k}})$ is expanded in powers of $m_{\rm P}^2$ in the subsequent way,
\be
S(q^A,v_{\vec{k}}) = m_{\rm P}^2\,S_0 + m_{\rm P}^0\,S_1 + m_{\rm
  P}^{-2}\,S_2 + \ldots \,. 
\ee
This ansatz is then inserted into \eqref{eq:WdWks} and for each power of $m_{\rm P}$, all the terms that are multiplied with a factor of $m_{\rm P}$ of that power are collected. The sum of terms with a specific power of $m_{\rm P}$ is then set equal to zero.

In the present case, the highest order that appears is $m_\mathrm{P}^4$, where we get the following equation
\be
\del{}{v_{\vec{k}}}\, S_0(q^A,v_{\vec{k}})=0\,,
\ee
which implies that the background part of the wave function represented by $S_0$ does not depend on the perturbations $v_{\vec{k}}$.

The next order is $m_\text{P}^2$, where we obtain the Hamilton--Jacobi equation of the background,
\be \label{HJeq}
\mathcal{G}_{AB}\,\del{S_0}{q_A}\,\del{S_0}{q_B}+V(q^A)=0\,.
\ee
It can be shown that this equation is equivalent to the Friedmann equation.

The subsequent order, which is $m_\mathrm{P}^0$, yields the equation
\begin{align} \label{eps0}
2\,\mathcal{G}_{AB}\,\del{S_0}{q_A}\,\del{S_1}{q_B} - \I\,\mathcal{G}_{AB}\,\del{^2 S_0}{q_A \partial q_B} &+ \left(\del{S_1}{v_{\vec{k}}}\right)^2 \nonumber\\
- \;\I\,\del{^2 S_1}{v_{\vec{k}}^2} &+ \omega^2_{\vec k}\,v_{\vec{k}}^2 = 0\,.
\end{align}
As one can see, the perturbations represented by $S_1$ and $v_{\vec{k}}$ now come into play. In order to obtain a Schr\"odinger equation for the perturbations modes, we define a wave function $\psi^{(0)}_{\vec{k}}$ in the following way
\be \label{def_psi0}
\psi^{(0)}_{\vec{k}}(q^A,v_{\vec{k}}) := \gamma(q^A)\,\E^{\I\,S_{1}(q^A,v_{\vec{k}})}\,.
\ee
Here we have introduced a prefactor $\gamma$, which is actually the inverse of the standard WKB prefactor that we did not include in our WKB-type ansatz \eqref{WKBansatz}. This is the procedure used in the general semiclassical expansion scheme for canonical quantum gravity (see, e.g., Eq.~(22) in Ref.~\cite{singh}). We demand that $\gamma$ obey the following condition:
\be \label{gammacond1}
\mathcal{G}_{AB}\,\del{}{q_A}\left[\frac{1}{2\gamma^2}\,\del{S_0}{q_B}\right] = 0\,.
\ee
Additionally, we can define the conformal WKB time, which we will identify with the classical conformal time, in terms of the minisuperspace variables as follows
\be \label{ctimedef}
\del{}{\eta} := \mathcal{G}_{AB}\,\del{S_0}{q_A}\,\del{}{q_B} = \E^{-2\alpha}\left[-\,\del{S_0}{\alpha}\del{}{\alpha}+\del{S_0}{\tilde{\phi}}\del{}{\tilde{\phi}}\right].
\ee
Using the last three relations, it is possible to rewrite equation \eqref{eps0} for $S_1$ as a Schr\"odinger equation
for $\psi^{(0)}_{\vec{k}}$:
\be \label{eq:Schreq}
\mathcal{H}_{\vec{k}} \psi^{(0)}_{\vec{k}}=\I\,\del{}{\eta}\,\psi^{(0)}_{\vec{k}}\,,
\ee
where the perturbation Hamiltonian $\mathcal{H}_{\vec{k}}$ is given by
\begin{align}
\mathcal{H}_{\vec{k}} :=-\,\frac{1}{2}\,\frac{\partial^2}{\partial v_{\vec k}^2}+\frac{1}{2}\,\omega^2_{\vec k}(\eta)\,v_{\vec{k}}^2\,. \label{SEqoneside}
\end{align}
Therefore, we end up with a Schr\"odinger equation for the quantum states of the
perturbations, where the time is defined from the minisuperspace quantum dynamics.

Up to now, we have recovered well-known physics, so the interesting part and main point of this investigation comes at the next order $m_\mathrm{P}^{-2}$, where quantum-gravitational corrections come into play. At this order, the equation we have to consider is given by
\begin{eqnarray} \label{eps-1}
\mathcal{G}_{AB}\,\del{S_0}{q_A}\,\del{S_2}{q_B} &+& \frac{1}{2}\,\mathcal{G}_{AB}\,\del{S_1}{q_A}\,\del{S_1}{q_B} - \frac{\I}{2}\,\mathcal{G}_{AB}\,\del{^2 S_1}{q_A \partial q_B} \notag\\
&+& \del{S_1}{v_{\vec{k}}}\,\del{S_2}{v_{\vec{k}}} - \frac{\I}{2}\,\del{^2 S_2}{v_{\vec{k}}^2} = 0\,.
\end{eqnarray}
The newly appearing function $S_2$, which contains information about the quantum-gravitational corrections at this order, is then split into a part $\varsigma$ that depends only on the minisuperspace variables, and a part $\chi$ that contains also the perturbations $v_{\vec{k}}$,
\be
S_2(q^A, v_{\vec{k}})\equiv\varsigma(q^A)+\chi(q^A,v_{\vec{k}})\,.
\ee
The reason for this split is to isolate the next-order correction to the WKB prefactor, which is given by $\varsigma$ and just contributes a phase. 
After this split, we end up with the subsequent equation for $\chi$:
\begin{eqnarray} \label{eta_dt}
\del{\chi}{\eta} = \frac{1}{\psi^{(0)}_{\vec{k}}}\Biggl(&-& \frac{1}{\gamma}\,\mathcal{G}_{AB}\,\del{\psi^{(0)}_{\vec{k}}}{q_A}\,\del{\gamma}{q_B} + \frac{1}{2}\,\mathcal{G}_{AB}\,\del{^2\psi^{(0)}_{\vec{k}}}{q_A \partial q_B} \notag\\
&+& \I\,\del{\psi^{(0)}_{\vec{k}}}{v_{\vec{k}}}\,\del{\chi}{v_{\vec{k}}} + \frac{\I\,\psi^{(0)}_{\vec{k}}}{2}\,\del{^2\chi}{v_{\vec{k}}^2}\Biggr).
\end{eqnarray}
We can now use $\chi$ to define a new wave function $\psi^{(1)}_{\vec{k}}$ that contains the quantum-gravitational corrections of the order $m_\mathrm{P}^{-2}$ as follows
\be \label{ansatz_corrWF}
\psi^{(1)}_{\vec{k}}(q^A,v_{\vec{k}}) := \psi^{(0)}_{\vec{k}}(q^A,v_{\vec{k}})\,\E^{\I\,m_\mathrm{P}^{-2}\,\chi(q^A,v_{\vec{k}})}\,.
\ee
Using this definition, we obtain a Schr\"odinger equation for $\psi^{(1)}_{\vec{k}}$ with a quantum-gravitational correction term that is suppressed by a prefactor of $m_\mathrm{P}^{-2}$
\begin{align} \label{corrSG_raw1}
\I\,\del{}{\eta}\,\psi^{(1)}_{\vec{k}}= \mathcal{H}_{\vec{k}}\psi^{(1)}_{\vec{k}} +\frac{\psi^{(1)}_{\vec{k}}}{m_\mathrm{P}^2\,\psi^{(0)}_{\vec{k}}}\Biggl(&\frac{1}{\gamma}\,\mathcal{G}_{AB}\,\del{\psi^{(0)}_{\vec{k}}}{q_A}\,\del{\gamma}{q_B} \notag\\
- \;&\frac{1}{2}\,\mathcal{G}_{AB}\,\del{^2\psi^{(0)}_{\vec{k}}}{q_A \partial q_B}\Biggr)\,.
\end{align}
We follow \cite{singh} in order to find an alternative expression for the
objects in the brackets in terms of the perturbation Hamiltonian
$\mathcal{H}_\vec{k}$. For illustrative purposes, we have outlined
this calculation in the appendix
 using only $\alpha$ instead of both minisuperspace variables. We finally obtain the following form of the quantum-gravitationally corrected Schr\"odinger equation,
\begin{align} \label{corrSchr}
\I\,\frac{\partial}{\partial \eta}\,\psi^{(1)}_{\vec{k}} = \mathcal{H}_{\vec{k}}\psi^{(1)}_{\vec{k}} -\frac{\psi^{(1)}_{\vec{\vec{k}}}}{2\,m_{\rm P}^2  \,\psi^{(0)}_{\vec{k}}}&\Biggl[\frac{\bigl(
\mathcal{H}_{\vec{k}}\bigr)^2}{V}\,\psi^{(0)}_{\vec{k}} \\
&+ \I\,\frac{\partial}{\partial \eta}\!
\left(\frac{\mathcal{H}_{\vec{\vec{k}}}}{V}\right)
\psi^{(0)}_{\vec{k}}\Biggr]\,. \nonumber
\end{align}
Note that, for a generic $\psi^{(0)}_{\vec{k}}$, the correction terms,
 which appear multiplying $\psi^{(1)}_{\vec{k}}$, are complex-valued.
 The imaginary component of those terms thus describes a potential source of
unitarity violation.

What is the source of this non-unitarity?
The crucial point is that the fundamental equation in quantum
cosmology is the Wheeler--DeWitt equation, not the Schr\"odinger
equation. First, the Hamiltonian in the Wheeler--DeWitt equation is
not self-adjoint, as is known for the general form of the quantum constraints in
geometrodynamics \cite{Komar}; self-adjointness is only
demanded for the Hamiltonian in the emerging (un-corrected)
Schr\"odinger equation \eqref{eq:Schreq}. Second, the Wheeler--DeWitt equation
obeys a Klein--Gordon type of conservation law, not the standard
Schr\"odinger conservation law. Expanding this Klein--Gordon law in
the Born--Oppenheimer scheme outlined above, one finds at order
$m_{\rm P}^0$ the Schr\"odinger conservation law, which at the next
order is {\em modified} by terms proportional to $m_{\rm P}^{-2}$ (see Ref.~\cite{honnef}, 
Eqs.~(3.12) ff.). In the present case, 
the expansion of the Klein--Gordon current leads at 
order $m_{\rm P}^{-2}$ to
\begin{equation}
\frac{\D}{\D\eta} \int\D v_{\vec{k}}\ \psi^{*(1)}_{\vec{k}}\psi^{(1)}_{\vec{k}}
= \frac{1}{m_{\rm P}^2}\int\D v_{\vec{k}}\ \psi^{*(1)}_{\vec{k}}
\frac{\partial}{\partial\eta}\left(\frac{\mathcal{H}_{\vec{\vec{k}}}}
{V}\right)\psi^{(1)}_{\vec{k}}.
\end{equation}
It is immediately recognizable that the term on the right-hand side
corresponds to the second term in the brackets on the right-hand side
of \eqref{corrSchr}. Imaginary terms such as the ones in
\eqref{corrSchr} have been applied before in the context of the
black-hole information problem \cite{KMS} and the quantum optics of
bosons in the gravitational field of the Earth \cite{lammerzahl}.

As we shall see in section \ref{sectcorr}, the non-unitary terms can become relevant
for early times. A consistent calculation of the quantum-gravitational
corrections to the power spectrum can, however, only be performed
if they are neglected there. This is similar to a process called
``unitarization'' that is frequently employed in particle physics
in order to avoid a non-unitary contribution to scattering amplitudes
from unknown high-energy physics; see e.g.~\cite{baak}, in particular Fig.~1-23 therein.

%We could in principle perform an appropriate
%redefinition of the wave function to absorb these unitarity-violating terms. However,
%as will be made explicit in the next section for a particular ansatz,
%it will only be possible to obtain a consistent and unambiguous
%framework by removing those imaginary terms.

\section{Gaussian ansatz}
\label{sectGauss}

\subsection{Gaussian ansatz for the Schr\"odinger equation} \label{sectGaussA}

Motivated by the fact that the most recent measurements of the CMB anisotropies have not shown any hint for a non-gaussianity of primordial fluctuations \cite{PlanckNG}, we assume that the scalar and tensor perturbations are in the ground state, which allows us to use a Gaussian ansatz to find a solution to the Schr\"odinger equation \eqref{eq:Schreq}. Introducing a normalization factor $N_{\vec{k}}^{(0)}(\eta)$ and the inverse Gaussian width $\Omega_{\vec{k}}^{(0)}(\eta)$, our ansatz reads:
\be
\label{Gaussianansatz2}
\psi^{(0)}_{{\vec{k}}}(\eta,v_{\vec{k}}) =
N_{\vec{k}}^{(0)}(\eta)\,\text{e}^{-\frac{1}{2}\,\Omega_{\vec{k}}^{(0)}(\eta)\,v_{\vec{k}}^2} .
\ee
We insert this ansatz into \eqref{eq:Schreq} and collect all the terms with either a factor of $v_{\vec{k}}^2$ or $v_{\vec{k}}^0$.
Each of the sums multiplying these terms is then set equal to zero and one ends up with the following two equations:
\begin{eqnarray}
\I\,N_{\vec{k}}^{(0)\prime}(\eta) &=&
\frac{1}{2}\,N_{\vec{k}}^{(0)}(\eta)\,
\Omega_{\vec{k}}^{(0)}(\eta)\,, \label{prefactor}
\\ 
\I\,\Omega_{\vec{k}}^{(0)\prime}(\eta) &=&
\bigl(\Omega_{\vec{k}}^{(0)}(\eta)\bigr)^2-\omega^2_{\vec k}(\eta)\,. \label{Omegazero}
\end{eqnarray}
In addition to these equations, we also request a normalized wave function
for any given time $\eta$:
\begin{align}
\left|\psi^{(0)}_{\vec{k}}\right|^2 &= \big|N_{\vec{k}}^{(0)}\big|^2\int\limits_{-\infty}^{\infty}\E^{-\frac{1}{2}\big[\Omega^{*(0)}_{\vec{k}}+\Omega_{\vec{k}}^{(0)}\big]v_{\vec{k}}^2}\,\D v_{\vec{k}} \notag
\\
& = \big|N_{\vec{k}}^{(0)}\big|^2 \frac{\sqrt{\pi}}{\sqrt{\Real\Omega_{\vec{k}}^{(0)}}} \;
\stackrel{!}{=}\; 1\,. \label{norm_N0}
\end{align}
This provides the modulus of the normalization factor in terms of
the real part of the inverse Gaussian width. Therefore, we get the following relation,
\be\label{normalization}
N_{\vec{k}}^{(0)}
=\left(\frac{\Real\Omega_{\vec{k}}^{(0)}}{\pi}\right)^{1/4}\, \E^{\I\varphi},
\ee
where $\varphi$ is a real function of time $\eta$. The imaginary
phase term would only be relevant in some kind of interference experiment,
but will not enter our results. It is easy to check that inserting this last relation into \eqref{prefactor} and separating it into its real and imaginary components,
yields two equations: an equation for the phase $\varphi$ and 
an equation for the real part of \eqref{Omegazero}.
Therefore, in order to obtain the complete wave function
(\ref{Gaussianansatz2}), up to a $v_{\vec{k}}$-independent phase term,
 it is enough to solve \eqref{Omegazero} for the inverse
Gaussian width. The normalization condition (\ref{normalization})
holds at any time and will ensure the correct normalization of the
wave function. 

Equation \eqref{Omegazero} is a Riccati equation and, as is well known,
it can always be written as a second-order linear differential equation.
With such a purpose, the following change of variable is performed,
\be
\label{Omega_k0}
\Omega_{\vec{k}}^{(0)}(\eta) = -\,\I\,\frac{y^{(0)\prime}_{\vec k}(\eta)}{y^{(0)}_{\vec k}(\eta)}\,,
\ee
which leads to the equation of a parametric oscillator,
\be
\label{yk}
y^{(0)\prime\prime}_{\vec k}(\eta) + \omega^2_{\vec k}(\eta)\,y^{(0)}_{\vec k}(\eta) = 0\,.
\ee
In fact, this equation corresponds to the equation of the classical perturbative variable
$v_{\vec k}$ or the quantized modes $\hat{y}_{\vec k}(\eta)$ in the Heisenberg picture
(see e.g.~\cite{BKKM99,PU09}).
There is, however, a subtlety. The functions
$y^{(0)}_{\vec k}(\eta)$ occurring in (\ref{Omega_k0}) are the {\em complex conjugates}
of the corresponding operators in the Heisenberg picture (see e.g.~\cite{CK92}).
This fact is very relevant when imposing initial conditions.

Let us be more specific. The standard assumption in the context of inflationary models
is to choose the Bunch--Davies vacuum at early times (or, equivalently, small scales)
$k\eta \rightarrow -\infty$. The physical interpretation is that the state
assumes the form of the Minkowski vacuum for large $k$. For a flat background,
the frequency $\omega$ is constant and equal to the wave number $k$. Therefore
it is straightforward to solve (\ref{yk}) as a linear combination of imaginary exponentials
with the mentioned frequency ($\E^{\pm {\rm i}k\eta}$). Now one just needs to choose
the positive frequency modes. In the Heisenberg picture, one would choose the solution
with a minus sign in the exponent but in this approach, because one has to take the
complex conjugate, the solution of the form $y^{(0)}_{\vec k}=\E^{\I k\eta}$ is the one that corresponds
to the Bunch--Davies vacuum.

Finally, the normalization of $y^{(0)}_{\vec k}(\eta)$ does not enter the physically relevant quantity
(\ref{Omega_k0}) and can be chosen as wanted. (Note that
\eqref{Omegazero} is of first order.)
 For convenience, we will follow the usual
convention in the Heisenberg picture and request the Wronskian to be equal to the imaginary
unit:
\be
\label{Wronskian}
W:= y^{(0)\prime}_{\vec k}y^{(0)*}_{\vec k} - y^{(0)\prime *}_{\vec k}y^{(0)}_{\vec k}=\I\,.
\ee

\subsection{Gaussian ansatz for the corrected Schr\"odinger equation} \label{sectGaussB}

Let us now study the same ansatz as the one presented in the previous subsection,
for the quantum-gravitationally corrected Schr\"odinger equation. Therefore, we write
the corrected wave function $\psi^{(1)}_{\vec k}$ in the following way,
\be
\label{Gaussianansatz3}
\psi^{(1)}_{{\vec{k}}}(\eta,v_{\vec{k}}) =
N^{(1)}_{\vec{k}}(\eta)\,\text{e}^{-\frac{1}{2}\,\Omega^{(1)}_{\vec{k}}(\eta)\,v_{\vec{k}}^2},
\ee
and insert this expression into \eqref{corrSchr}. As before, different orders of $v_{\vec{k}}$
are set independently to zero. At order $v_{\vec{k}}^4$, the following algebraic relation is obtained,
\begin{equation}
N^{(1)}_{\vec{k}} \left[\omega_{\vec k}^2-(\Omega^{(0)}_{\vec{k}})^2\right]=0.
\end{equation}
This equation can obviously not be obeyed by any non-trivial wave function. Therefore, since it
is of higher order in the perturbations $v_{\vec{k}}$, we neglect it. Apart from this equation,
at orders $v_{\vec{k}}^0$ and $v_{\vec{k}}^2$ equations similar to (\ref{prefactor}) and (\ref{Omegazero})
are respectively obtained, but with correction terms. In particular the equation
for the inverse Gaussian width takes the following form:
\begin{eqnarray}\label{Omegaeq}
\I\,\Omega^{(1)\prime}_{\vec{k}} =(\Omega^{(1)}_{\vec{k}})^2-\widetilde\omega_{\vec{k}}^2\,,
\end{eqnarray}
where the corrected frequency $\widetilde\omega_{\vec{k}}$ is defined in the following way,
\begin{align}\label{defcorrectedomega}
\widetilde\omega_{\vec{k}}^2:=\omega_{\vec{k}}^2-
\frac{1}{2 m_\mathrm{P}^2 V}
\biggl[&\left(3\Omega^{(0)}_{\vec{k}}-{\rm i}\,(\ln V)'\right) \left(\omega^2_{\vec{k}} - (\Omega^{(0)}_{\vec{k}})^2\right) \notag \\
&+2\, \I \,\omega_{\vec{k}} \, \omega'_{\vec{k}}\biggr].
\end{align}
The differential structure of the equation is the same and thus
one can perform the same change of variable as above
in \eqref{Omega_k0}, that is,
\be
\label{Omega_k}
\Omega^{(1)}_{\vec{k}}(\eta) = -\,\I\,\frac{y^{(1)\prime}_{\vec k}(\eta)}{y^{(1)}_{\vec k}(\eta)}\,,
\ee
to obtain a linear equation for the auxiliary variable $y^{(1)}_{\vec k}$,
\be
\label{ykcorrected}
y^{(1)\prime\prime}_{\vec k}(\eta) + \widetilde\omega^2_{\vec k}(\eta)\,y^{(1)}_{\vec k}(\eta) = 0\,.
\ee
This last equation is quite complicated and, as will be shown below
in the application of the formalism to a de Sitter background, it will not
be easy to obtain analytical solutions for it. Therefore, one could deal with
this equation numerically (this will be performed in section \ref{sectcorrfull}) or one
could also consider its linearization around $\Omega^{(0)}_{\vec k}$.
In this last case,
we write the inverse Gaussian width as
\be
\Omega^{(1)}_{\vec k}=\Omega^{(0)}_{\vec k}+\widetilde\Omega^{(1)}_{\vec k}\,.
\ee
If the function $\widetilde\Omega^{(1)}_{\vec k}$ is assumed to be small, one can drop its quadratic terms.
In this way, the following equation is obtained for $\widetilde\Omega^{(1)}_{\vec k}$,
\be\label{linearizedeqomega}
{\rm i}\, \widetilde\Omega^{(1)\prime}_{\vec k}=2\, \Omega^{(0)}_{\vec k} \widetilde\Omega^{(1)}_{\vec k}-\left(\widetilde\omega^2_{\vec k}-\omega^2_{\vec k}\right),
\ee
which corresponds to the differential equation for the equivalent to $\widetilde\Omega^{(1)}_{\vec k}$ studied in \cite{KK-PRL,BEKKP}.

Up to this point, everything seems to be fine. But we have not yet considered
the equation for the normalization factor $N^{(1)}_{\vec k}$. In fact, the problem appears
when one wants to impose a normalization condition.
Proceeding in the same way as we did in the previous subsection, 
and making thus use of the usual inner product of quantum mechanics since
the evolution equation is the Schr\"odinger equation plus certain linear terms
in $\psi^{(1)}_{\vec k}$, one would request the same relation \eqref{norm_N0}
that provides the normalization factor in terms of the inverse Gaussian width.
The issue is that, when this relation is inserted into the evolution equation
for $N^{(1)}_{\vec k}$, an equation is obtained for the real part of the Gaussian width
$\Omega^{(1)}_{\vec k}$ that is in conflict with \eqref{Omegaeq}. This is a signal
of nonunitarity and the system is telling us that the normalization of the
wave function cannot be conserved. This conflict is due to the presence
of imaginary terms in the factor that multiplies $\psi^{(1)}_{\vec k}$ in the second
term of the right-hand side of the corrected Schr\"odinger equation (\ref{corrSchr}).
The only way to get a consistent framework, where one can unambiguously
take expectation values, which is certainly necessary to compute the power
spectrum, is to get rid of those imaginary terms. In the present Gaussian
ansatz, this would translate to taking only the real part of the corrected squared frequency
$\widetilde\omega_{\vec k}^2$ (\ref{defcorrectedomega}) in all equations of this section,
specifically in (\ref{Omegaeq}), (\ref{ykcorrected}) and (\ref{linearizedeqomega}).

In addition, in section \ref{sectcorrfull} we will analyze numerically, for a de Sitter background, the solutions of (\ref{Omegaeq})
for both cases: on the one hand, with the complete corrected frequency $\widetilde\omega_{\vec k}$
and, on the other hand, with only its real part. There, it will be made explicit that the solution with the complete frequency
has more undesirable physical properties. In particular, it is clear that the solution to (\ref{ykcorrected}) with a real frequency will be composed of oscillatory functions. But having imaginary terms in the frequency will induce exponential-like solutions, which will generically imply, as will be explicitly shown below, an amplification of the amplitude of the wave either towards the past or towards the future.

\section{Derivation of the power spectra} \label{sectPS}

In order to obtain the power spectra of the scalar and tensor perturbations (cf.~e.\,g.~\cite{Martin:2012}), we start from the two-point correlation function of the corresponding master variable $v$, which is given by
\begin{eqnarray}
\Xi(\vec{r}) &:=&\left \langle \psi_{\vec k}\left \vert \hat{v}(\eta ,{\vec x}) 
\hat{v}(\eta,{\vec x}+{\vec r})\right \vert \psi_{\vec k}\right \rangle \\
&=&\int \prod _{\vec k}{\rm d}v_{\vec k}\,
\psi_{\vec k}^*(v_{\vec k})\,
v(\eta ,{\vec x})\,
v(\eta,{\vec x}+{\vec r})\,
\psi_{\vec k}(v_{\vec k})\,. \notag
\end{eqnarray}
Using the Gaussian ansatz (\ref{Gaussianansatz2}), this expression can be rewritten as
\be
\Xi(r)
= \frac{1}{2\pi^2}\int_0^{+\infty}
\frac{{\rm d}p}{p}\,\frac{\sin(pr)}{pr}\,p^3\,\frac{1}{2 \,\Real  \Omega_{\vec p}}\,; 
\ee
the power spectrum for $v$ is then defined as
\be\label{pev}
{\cal P}_v(k)=\frac{k^3}{4\pi^2}\,\frac{1}{\Real  \Omega_{\vec k}}\,.
\ee
This formula is valid for both cases presented in the previous section with
or without considering quantum-gravitational corrections. In the case without
corrections, one would just replace $\Omega_{\vec k}$ by $\Omega^{(0)}_{\vec k}$ in the last expression.
As can be explicitly seen in this
last formula, the real part of the inverse Gaussian width is the object
that contains the complete information about the corresponding power
spectrum.

As usual, the two-point correlation function of the Fourier-transformed variable $v_{\vec k}$
can also be written in terms of this power spectrum:
\be
\label{eq:twopointfourier}
\left \langle \psi_{\vec k}\left \vert \hat{v}_{\vec k}
\hat{v}_{\vec p}^*\right \vert \psi_{\vec k} \right \rangle 
=\frac{2\pi^2}{k^3}\,{\cal P}_v(k)\,\delta({\vec k}-{\vec p})\,.
\ee
In order to explicitly write out the power spectrum corresponding to the scalar or
tensor modes, respectively, one has to define which is the physically relevant variable
in each case. This will be done in the following.

\subsection{Power spectrum for scalar perturbations}

For the case of the scalar perturbations, the physical quantity we are interested in are the temperature anisotropies of the CMB, which are 
related to the comoving curvature perturbation $\zeta$ defined by
\be \label{defzeta}
  \zeta:=\Phi_{\rm B}+\frac{2}{3}\frac{\Phi_{\rm B}'
+{\mathscr{H}}\Phi_{\rm B}}{{\mathscr{H}}(1+ w)}\,.
\ee
Here, $\Phi_{\rm B}$ represents the Bardeen potential (\ref{eq:defbardeen}),
and $w$ stands for the barotropic index appearing in the equation of state $P=w\rho$. 
This expression can be simplified for our purposes, because the CMB anisotropies are created during the matter-dominated era, where $w=0$. Furthermore, taking into account only large-scale perturbations, we can regard $\Phi_{\rm B}$ as approximately constant and end up with (cf.~also e.g.~\cite{Martin:2012})
\be
\zeta\simeq \frac{5}{3}\,\Phi_{\rm B}\,.
\ee
We now have to express $\zeta$ in terms of the master perturbation variable $v_{\vec k}$ we use. The relation is given by
\be
\label{eq:linkzetav}
\zeta_{\vec k} =\sqrt{\frac{4\pi G}{\epsilon}}\,\frac{v_{\vec k}}{a}\,,
\ee
where the slow-roll parameter $\epsilon$ has already been defined in \eqref{defepsilon}.

The power spectrum for the scalar perturbations is thus given by the power spectrum of the curvature perturbations $\zeta_{\vec k}$ and using (\ref{pev}) we can finally write
\be \label{PS0omega}
{\cal P}_{\rm S}(k):={\cal P}_{\zeta}(k)=
 \frac{4\pi G}{a^2\,\epsilon}\,\frac{k^3}{2\pi^2}\,\frac{1}{2\,\Real \Omega_{\vec k}}\,.
\ee
It is important to note that, before $\Real  \Omega_{\vec k}(\eta)$
is inserted into the last expression,
one has to take the limit of super-Hubble scales (or late times) given by $k\eta \rightarrow 0^-$.
As we will see, the perturbations get ``frozen" in this limit.

\subsection{Power spectrum for tensor perturbations}

For the case of the tensor perturbations, the power spectrum for the master perturbation variable $v_{\vec k}$ also takes the form
(\ref{pev}), but instead of the curvature perturbations $\zeta_{\vec k}$, one now has to evaluate the following variable
\be
h_{\vec k} = 2\sqrt{8\pi G}\,\frac{v_\vec{k}}{a}\,.
\ee
Given that the tensor modes have two polarizations, the power spectrum corresponding to the variable $h$ has to be multiplied by a factor of two, such that we obtain the subsequent power spectrum for the tensor perturbations,
\be
\label{PT0omega}
{\cal P}_{\text{T}}(k) := 2\,{\cal P}_{h}(k)= \,\frac{64\pi G}{a^2}\,\frac{k^3}{2\pi^2}\,\frac{1}{2\,\Real \Omega_{\vec k}}\,.
\ee
Here, the same comment as in the scalar case about taking the super-Hubble scale limit for $\Real \Omega_{\vec k}$ applies.

Finally, we also define the tensor-to-scalar ratio $r$ as
\be \label{tts0}
r := \frac{{\cal P}_{\text{T}}(k)}{{\cal P}_{\text{S}}(k)}\,.
\ee
This is the relevant dimensionless quantity that
expresses the relative strength of the primordial gravitational wave
background. 

\section{The uncorrected spectra for a de~Sitter background} \label{sectdS}

In this section, we solve the uncorrected Schr\"odinger equation and use it to derive the
power spectra for scalar and tensor modes in a de Sitter background.
In this way, we will find the standard result from a different perspective.
The quantum-gravitational correction to this power spectrum will then be
derived in the next section. Let us comment that, to be precise, pure de Sitter
is a singular limit of the formalism developed in the previous sections.
The background matter field will be fixed to a constant value and, thus,
in principle it does not make sense to speak about its (scalar) perturbations.
This is shown explicitly in the appearance of the slow-roll parameter $\epsilon$,
which in this limit should be vanishing, in \eqref{defz} and \eqref{eq:linkzetav},
as well as in the divergence of the curvature perturbation $\zeta$ \eqref{defzeta},
since for pure de Sitter $w=-1$. In fact, in this case one actually
does not obtain a power spectrum since the modes never re-enter
the horizon. Nonetheless, the computations can be performed by keeping
the slow-roll parameter as a non-vanishing constant, and this particular case
represents a very good and easy-to-compute example to test the developed formalism.

Therefore, as commented above, we construct the de Sitter universe by setting the scalar field $\phi$ to a constant value.
Using the vanishing of the background Hamiltonian \eqref{backgroundH}, and neglecting the
time derivative of $\phi$, we can write the following relation between the potential $\mathcal{V}$
of the scalar field and the constant Hubble parameter $H_0$ in our de Sitter universe,
\be \label{Vmp}
\mathcal{V}(\phi) = \frac{3}{8\pi G}\,H_0^2 = \frac{1}{2}\,m_\text{P}^2\,H_0^2\,.
\ee
In order to obtain a concrete relation between $\phi$ and $H_0$, one has to choose a specific potential, e.g.~$\mathcal{V}(\phi) = \frac{1}{2}\,m^2\phi^2$, such that $\phi = m_\text{P}H_0/m$, but we will refrain from doing this in order to keep our considerations as general as possible.

The auxiliary potential $V(\eta)$ defined in \eqref{auxV} thus reads
\be
V(\eta) = \E^{4\alpha}H_0^2 = \frac{1}{H_0^2\eta^4}\,.
\ee
In the de Sitter case, the frequencies $\omega^2_{\vec k}(\eta)$
take the same form for both scalar and tensor perturbations,
\be \label{omkdS}
\omega^2_{\vec k}(\eta) = k^2 - \frac{2}{\eta^2}\,,
\ee
such that we can consider both kinds of perturbations in one step up to the calculation of the power spectra.

Now we can insert these relations into the equations obtained in the semiclassical approximation presented in section \ref{sectsemi} and from the Gaussian ansatz applied in section \ref{sectGauss}.

Setting $\phi$ to a constant value implies that we have to neglect the derivative with respect to $\phi$ in the Hamilton--Jacobi equation \eqref{HJeq}, which thus simplifies to
\be
\left(\frac{\partial S_0}{\partial\alpha}\right)^2 
- \E^{6\alpha}H_0^2 =0\,.
\ee
Its solution can be easily obtained and reads
\be \label{solHJ}
S_0(\alpha)= -\,\frac{1}{3}\,\E^{3\alpha}H_0\,,
\ee
where we have chosen the minus sign, which corresponds to an expanding universe. This is then also reflected in the definition of the WKB conformal time \eqref{ctimedef}, given here by
\be
\del{}{\eta} = \E^{\alpha}H_0\,\del{}{\alpha}\,.
\ee
This choice recovers the standard flux of time, that is, $\eta$ growing with $\alpha$, and thus can be directly
identified with the classical conformal time.

Considering the derivations presented in sections \ref{sectGaussA} and \ref{sectPS},
we have seen that in order to obtain the power spectra, we just need to find a solution for $\Omega^{(0)}_{\vec k}$
from equation \eqref{Omegazero}, which, for this particular case, is given by
\be \label{dglOm0}
\I\,\Omega_{{\vec k}}^{(0)\prime}(\eta) =
\bigl(\Omega_{{\vec k}}^{(0)}(\eta)\bigr)^2-k^2 + \frac{2}{\eta^2}\,.
\ee
Then, its real part $\Real\Omega^{(0)}_{\vec k}$, in the limit of super Hubble scales
($\eta\rightarrow 0$), should be inserted into expressions (\ref{PS0omega}) and \eqref{PT0omega}.
In fact,
equation \eqref{dglOm0} can be easily solved via the auxiliary variable $y^{(0)}_{\vec k}$ defined in \eqref{Omega_k0} and its solution is
given by particular Hankel functions which can be expressed in 
terms of elementary functions (see e.g.~\cite{PU09}, p.~472) as follows,
\be
\label{yk-solution}
y^{(0)}_{\vec k}(\eta) = A(k)\,\E^{-\I k\eta}\left(1-\frac{\I}{k\eta}\right)
+B(k)\,\E^{\I k\eta}\left(1+\frac{\I}{k\eta}\right).
\ee
The Bunch--Davies vacuum is imposed as initial state, which, as explained at the end of section \ref{sectGaussA},  for this solution means choosing $A(k)=0$. This leads to 
\be
\label{BD}
y^{(0)}_{\vec k}(\eta) = 
\frac{1}{\sqrt{2k}}\,\E^{\I k\eta}\left(1+\frac{\I}{k\eta}\right),
\ee
where the normalization has been chosen such that the corresponding Wronskian is equal
to the imaginary unity. Consequently, we find from (\ref{dglOm0})
\be \label{om0eta}
\Omega_{{\vec k}}^{(0)}(\eta)=\frac{k^3\eta^2}{1+k^2\eta^2}+\frac{{\rm i}}{\eta(1+k^2\eta^2)}\,.
\ee
From (\ref{om0eta}), we see that $\Omega_{{\vec k}}^{(0)}(\eta)$ approaches $k$ for 
$\eta\to-\infty$. Note that this corresponds, of course, to the expression for the
Minkowski vacuum and justifies the choice of (\ref{BD}) in this Schr\"odinger picture. 
If we had chosen, for example, $B(k)=0$ in (\ref{yk-solution}), we would have found
in this limit $\Omega_{{\vec k}}^{(0)}(\eta)=-k$, which would lead to a non-normalizable
Gaussian, which is physically not acceptable. 

As explained above, before plugging the real part of the solution \eqref{om0eta} into the expressions for the power spectra \eqref{PS0omega} and \eqref{PT0omega}, one needs to take the large-scale limit $k\eta \rightarrow 0^-$, which leads to
\be \label{om0shor}
\Real\Omega_{{\vec k}}^{(0)}(\eta) = \frac{k^3\eta^2}{1+k^2\eta^2} \; \longrightarrow \; k^3\eta^2\,.
\ee
If we furthermore use the relation between scale factor and conformal time,
\be
a = -\,\frac{1}{H_0\eta}\,,
\ee
we obtain from \eqref{PS0omega} the following power spectrum for the
scalar modes: 
\be \label{Ps0}
{\cal P}_{\text{S}}^{(0)}(k) 
= \frac{G\,H_0^2}{\pi\epsilon}\biggr|_{k=H_0a}\,.
\ee
Given that the expression for the scalar power spectrum contains the slow-roll parameter $\epsilon$, we have to evaluate \eqref{Ps0} at that point in time 
at which a certain mode $k$ reenters the Hubble scale, which is given by $k=H_0 a$.  This makes the power spectrum \eqref{Ps0} become slightly $k$-dependent.
This is, of course, the standard result (see e.g.~\cite{PU09}, Sec.~8.4), but derived here
in the Schr\"odinger picture, starting from an expansion of the quantum-gravitational
Wheeler--DeWitt equation.

Following the same procedure for the tensor modes, we obtain their corresponding
power spectrum from \eqref{PT0omega},
\be
\label{PT0}
{\cal P}_{\text{T}}^{(0)}(k) = \frac{16 G\,H_0^2}{\pi}\,.
\ee
Hence, we also recover the standard result for the tensor-to-scalar ratio $r$,
\be \label{tts0}
r^{(0)} = \frac{{\cal P}_{\text{T}}^{(0)}(k)}{{\cal P}_{\text{S}}^{(0)}(k)} =  16\, \epsilon\,.
\ee
If we re-insert $\hbar$ and $c$ in the last results, we can write
\begin{equation*}
{\cal P}_{\text{S}}^{(0)}(k) = \frac{(t_{\rm P}H_0)^2}{\pi\epsilon}\biggr|_{k=H_0a},\quad
{\cal P}_{\text{T}}^{(0)}(k) =\frac{16(t_{\rm P}H_0)^2}{\pi}\,,
\end{equation*}
where $t_{\rm P}$ is the Planck time. This expresses the fact that already
(\ref{Ps0}) is a quantum-gravitational result (see also \cite{Woodard}). The reason is that the scalar
perturbations in the metric together with the inflaton field are combined
into a variable, the Mukhanov--Sasaki variable $v$, which is
quantized. In the limit 
$\hbar\to 0$, no effect is obtained. The quantum-gravitational
corrections that will be addressed 
in the next section correspond, in the usual language, to one-loop
quantum corrections  
\cite{BK98}, which arise from the fact that with our main
Wheeler--DeWitt equation \eqref{eq:WdWmaster}, we consider full
quantum gravity instead of just a quantum-mechanically perturbed
spacetime as in the conventional approach to cosmological
perturbations. 

\section{The corrected spectra for a de~Sitter background}
\label{sectcorr}

Let us now finally calculate the quantum-gravitational corrections originating from \eqref{corrSchr} to the power spectra obtained in the previous section for a de Sitter universe. As we have derived in section \ref{sectGaussB}, for this purpose, we need to find a solution to the differential equation \eqref{Omegaeq} or its linearized version \eqref{linearizedeqomega}. This is the goal of this section. More precisely, in the first two subsections we
will present the analysis of the full equation mostly by numerical methods. In
a third subsection, we will deal with the linearized version, which allows for
an analytical solution.

\subsection{Numerical analysis of the full equation: Bunch--Davies initial conditions}
\label{sectcorrfull}

The aim of this subsection is to analyze the generic behavior of the solution of the full equation \eqref{Omegaeq}, with and without
the imaginary (unitarity-violating) terms in $\widetilde\omega_{{\vec k}}^2$. For such purpose, Bunch--Davies initial data will be considered.
In the subsequent subsection, we will discuss modified, more natural, initial conditions for the corrected Schr\"odinger equation.

\begin{figure}
\includegraphics[width=0.5\textwidth]{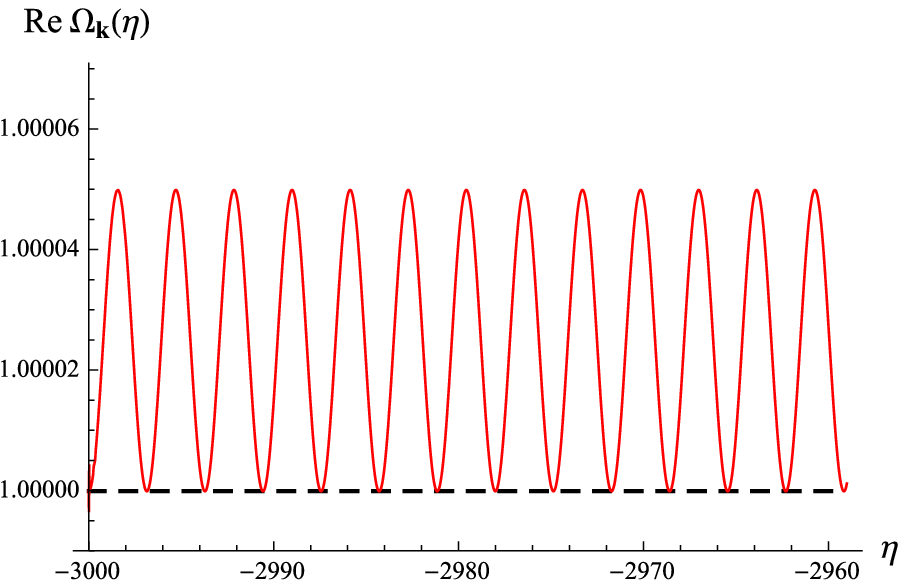} \\
\vspace{0.5cm}
\includegraphics[width=0.5\textwidth]{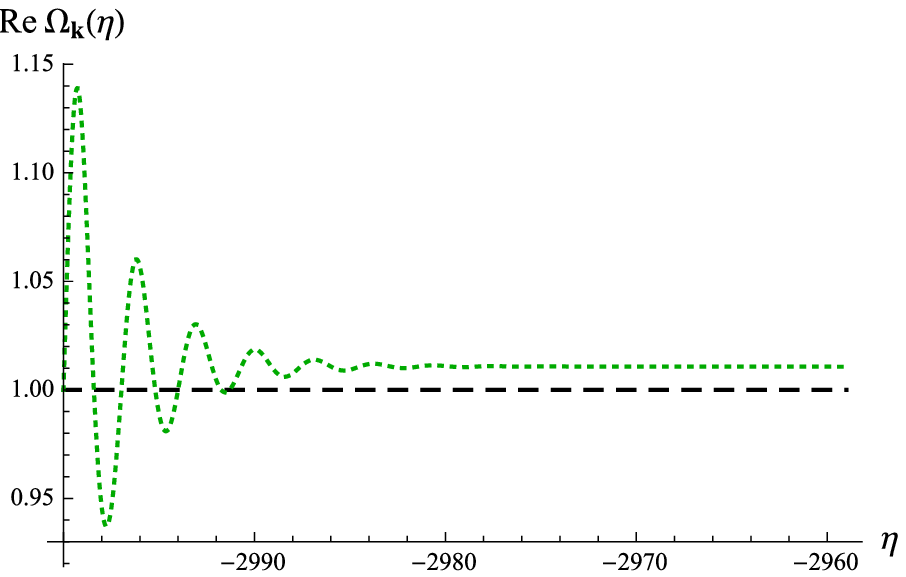} \\
\vspace{0.5cm}
\includegraphics[width=0.5\textwidth]{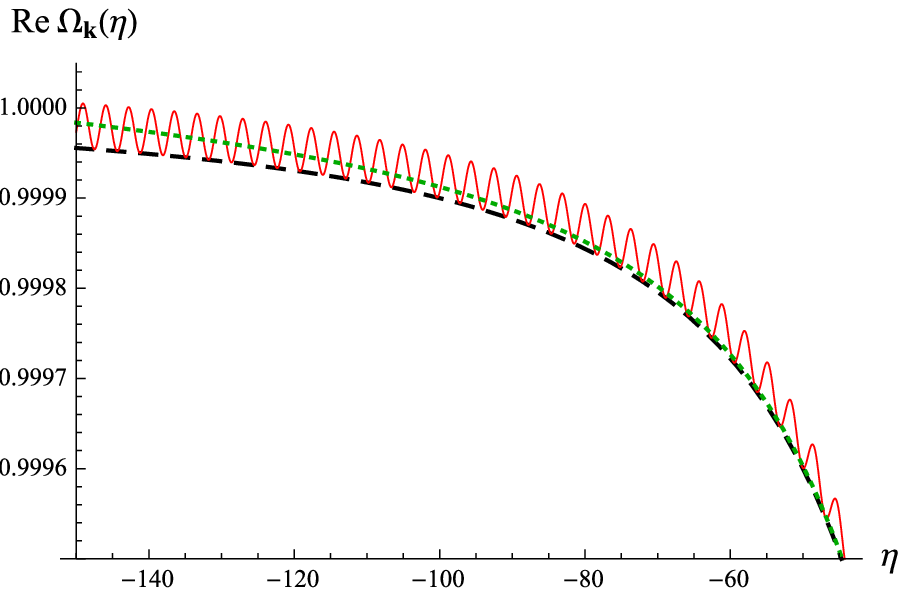}
\caption{The real parts of  $\Omega_{{\vec k}}^{(0)}$ in black (dashed line),
of $\Omega_{{\vec k}}^{(1)}$ with its complete equation of motion in green (dotted line),
and of $\Omega_{{\vec k}}^{(1)}$ removing the imaginary terms of the frequency from its
equation in red (continuous line) are shown for $k=1$. The first two plots correspond to early
times. In the first plot one can see the constant amplitude oscillations of the
solution without the imaginary terms, whereas in the second plot the
damped oscillations of the solution with imaginary terms is clearly seen.
The last plot corresponds to late times.}\label{omegaearly}
\end{figure}

It is very difficult to find an analytical solution for the differential equation \eqref{Omegaeq}, which is given by
\be\label{Omegaeqdesitter}
\I\,\Omega'^{(1)}_{\vec{k}} =(\Omega^{(1)}_{\vec{k}})^2-\widetilde\omega_{\vec{k}}^2\,,
\ee
where, for the de Sitter case,  the corrected frequency \eqref{defcorrectedomega} reads explicitly as follows,
\begin{align}
\widetilde\omega_{{\vec k}}^2=\omega_{{\vec k}}^2-
\frac{H_0^2\eta^4}{2 m_\mathrm{P}^2}
\biggl[&\left(3\Omega^{(0)}_{{\vec k}}+\frac{4{\rm i}}{\eta}\right) \left(\omega^2_{{\vec k}} - (\Omega^{(0)}_{{\vec k}})^2\right) \notag \\
&+2\, \I \,\omega_{{\vec k}} \, \omega'_{{\vec k}}\biggr].
\end{align}
Therefore, we resort to numerical methods, and, as commented above, we shall consider both the equation with the complete $\widetilde\omega_{{\vec k}}^2$ as well as with only the real part of $\widetilde\omega_{{\vec k}}^2$. In the latter case, the corrected frequency to be considered simplifies to
\be \label{realomtilde}
\Real\!\bigl(\widetilde\omega_{{\vec k}}^2\bigr) = \omega_{{\vec k}}^2 -\frac{H_0^2\eta^4}{2 m_\mathrm{P}^2}\,\frac{k^3(11-k^2\eta^2)}{(1+k^2\eta^2)^3}\,.
\ee
Numerically, the Bunch--Davies initial conditions are just implemented by
requesting $\Omega^{(1)}_{\vec k}$ to be equal to 
the exact solution \eqref{om0eta} we have found in the previous section at a chosen initial time.
In the particular case of the plots shown in Fig.~\ref{omegaearly}, we have chosen an initial value of $\eta_{\rm initial}=-3000$ and set the ratio to the unrealistically large value $H_0/m_\mathrm{P}=10^{-2}$ such that the corrections can be seen in the plots. We then solve the equation \eqref{Omegaeqdesitter}
numerically both with the complete $\widetilde\omega^2_{{\vec k}}$ and with its imaginary part dropped.

For the case of the equation with the complete $\widetilde\omega_{\vec k}^2$, 
we find that in the limit of large (negative) $\eta$, $\Real\Omega_{{\vec k}}^{(1)}$ shows damped oscillations towards the future and, therefore, amplified towards $\eta=-\infty$. As already commented above, this is a consequence expected from
 having a complex frequency, which is another argument against keeping the unitarity-violating terms.

On the other hand, for the case of the equation with only the real part of $\widetilde\omega_{\vec k}^2$,
 $\Real\Omega_{{\vec k}}^{(1)}$ turns out to be an oscillating function with a constant amplitude.
 More precisely, it oscillates around
\be
k+\frac{H_0^2}{4 k\,m_\text{P}^2}\,,
\ee
with an amplitude of $H_0^2/(4 k m_\text{P}^2)$.
In this way, the minima of $\Real\Omega_{{\vec k}}^{(1)}$ coincide with the value of $\Real\Omega_{{\vec k}}^{(0)}$, which at early times is equal to $k$. This oscillatory behavior, with a constant amplitude and a minimum given by $\Real\Omega_{{\vec k}}^{(0)}$ is obeyed all along until very small values of $\eta$
(see Fig.~\ref{omegaearly}). In fact, for large negative values of $\eta$, the explained
behavior could be expected beforehand due to the specific initial data we requested,
which is to have an almost vanishing derivative at initial time at a value equal to $\Omega_{{\vec k}}^{(1)}$.
Requesting an almost zero time derivative at the initial point imposes that such point should be a
minimum and thus fixes the amplitude of the oscillations. Therefore, this amplitude should, in principle,
not be regarded as physically meaningful since we are imposing it by hand.

As for the behavior of the numerical solution for $\eta \rightarrow 0^-$, we find that both solutions follow the same
qualitative behavior as $\Real\Omega_{{\vec k}}^{(0)}$, but they differ quantitatively, which implies a correction in the spectrum.
In subsection \ref{linearizedsection}, by making use of the linearized version of
the equation for $\Omega^{(1)}_{{\vec k}}$, we will compute quantitatively such a correction.

\subsection{Numerical analysis of the full equations: natural initial conditions}

\begin{figure}[h]
\includegraphics[width=0.5\textwidth]{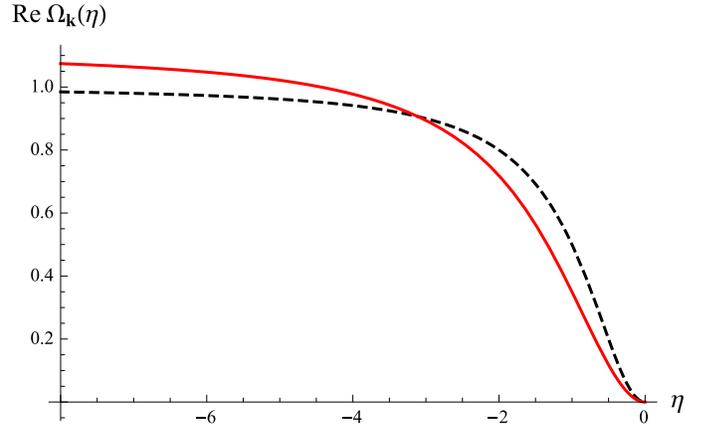}
\caption{The real part of $\Omega^{(1)}_{\vec{k}}$ with Bunch--Davies-like initial data (red continuous line) compared to $\Real\Omega^{(0)}_{\vec{k}}$ (black dashed line) for $k = 1$ with the unrealistically high ratio $H_0/m_\text{P} = 2/3$.
}
\label{plotom01lin}
\end{figure}

\begin{figure}[h]
\includegraphics[width=0.5\textwidth]{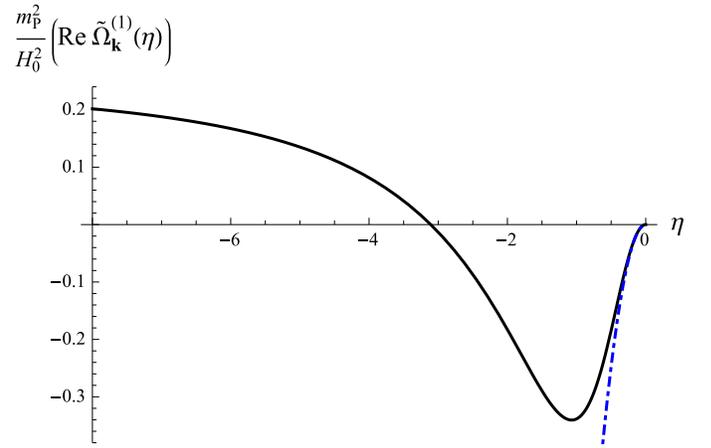}
\caption{The real part of $\widetilde\Omega^{(1)}_{\vec k}:=\Omega^{(1)}_{\vec k}-\Omega^{(0)}_{\vec k}$ (black solid line) and its expansion \eqref{om1linlimit0}, with the integration constant $c_1=0$,
at $\eta \rightarrow 0$ (blue dot-dashed line) for $k=1$.}
\label{plottildeom1lin}
\end{figure}

As has been commented above, the behavior of the $\Omega^{(1)}_{{\vec k}}$ will be highly dependent on the chosen
initial conditions. Therefore, one should argue which would be its natural initial data.
In the case of  $\Omega^{(0)}_{{\vec k}}$ the usual Bunch--Davies vacuum is regarded as the initial physically meaningful state.
The idea behind the Bunch--Davies vacuum is that at the beginning, when each mode is well inside its
Hubble horizon, it is oscillating with a constant frequency $k$. This means that $y^{(0)}_{\vec{k}}=\E^{\I k\eta}$
and therefore $\Omega^{(0)}_{{\vec k}}=k$. Following the same argumentation, in the corrected case, we should
try to find a frequency that makes the real part of $\Omega^{(1)}_{{\vec k}}$ equal to a constant $\beta_{\vec k}$
(which would stand for the frequency of the free modes), whereas its imaginary part is vanishing.
In this way, we would find that the mode function would be a freely oscillating function $y^{(1)}_{\vec{k}}=\E^{\I \beta_{\vec k}\eta}$.
By writing explicitly the equation of motion for $\Real(\Omega^{(1)}_{{\vec k}})$ and
$\Imag(\Omega^{(1)}_{{\vec k}})$ from (\ref{Omegaeq}),
\begin{eqnarray*}
\Real(\Omega^{(1)}_{{\vec k}})'&=& 2\,\Real(\Omega^{(1)}_{{\vec k}})\,\Imag(\Omega^{(1)}_{{\vec k}})-\Imag(\widetilde\omega^2_{\vec k})\,,\\
\Imag(\Omega^{(1)}_{{\vec k}})'&=& \Imag(\Omega^{(1)}_{{\vec k}})^2-\Real(\Omega^{(1)}_{{\vec k}})^2+\Real(\widetilde\omega^2_{\vec k})\,,
\end{eqnarray*}
it is easy to see that
this can only be achieved if the imaginary part of $\widetilde\omega_{\vec k}$ (the unitarity-violating part) is
vanishing and initially ($\eta\rightarrow-\infty$) $\Real(\Omega^{(1)}_{{\vec k}})^2=\Real(\widetilde\omega^2_{{\vec k}})$.
From the explicit form of the corrected frequencies given in \eqref{realomtilde}, we can compute the value of its real part
at an initial time $\eta\rightarrow-\infty$,
\be
\beta_{\vec k}:=\sqrt{\Real(\widetilde\omega_{\vec k}^2)}=\sqrt{k^2+\frac{H_0^2}{2 k\,m_\text{P}^2}}\approx k+\frac{H_0^2}{4 k\,m_\text{P}^2}\,,
\ee
which is exactly the mean value of the oscillations found in the numerical analysis above.
Therefore, these latter conditions ($y_{\vec
  k}^{(1)}=\E^{\I\beta_\vec{k}\eta}$ or, equivalently,
$\Real(\Omega^{(1)}_{\vec k})=\beta_\vec{k}$ and
$\Imag(\Omega^{(1)}_{\vec k})=0$) would be the natural
(Bunch--Davies-like) initial conditions that one should consider when
solving the corrected equation of motion. 
With such conditions one gets a non-oscillating evolution for $\Real\Omega_{{\vec k}}^{(1)}$.
The plot in Fig.~\ref{plotom01lin} shows this non-oscillating solution as compared to the real part of $\Omega_{{\vec k}}^{(0)}$.
In addition, in
Fig.~\ref{plottildeom1lin} we plot the difference $\Real\widetilde\Omega_{{\vec k}}^{(1)}$ between this
non-oscillating numerical solution $\Real\Omega_{{\vec k}}^{(1)}$ and $\Real\Omega_{{\vec k}}^{(0)}$ for late times.
In the following section we will also make use of these initial conditions to fix the integration constant
that will appear when solving the linearized equation for $\widetilde\Omega_{{\vec k}}^{(1)}$.

\subsection{Analysis of the linearized equation}
\label{linearizedsection}

Since the numerical solution we have discussed above does not allow us to determine an analytical
behavior of the quantum-gravitational corrections, we now turn to the linearized equation for $\widetilde\Omega^{(1)}_{\vec{k}} = \Omega^{(1)}_{\vec k}  - \Omega^{(0)}_{\vec k}$ as given in \eqref{linearizedeqomega}. This means, we will solve the equation
\be \label{dglom1}
{\rm i}\, \widetilde\Omega^{(1)\prime}_{\vec k}=2\, \Omega^{(0)}_{\vec k} \widetilde\Omega^{(1)}_{\vec k}-\left(\widetilde\omega^2_{\vec k}-\omega^2_{\vec k}\right),
\ee
with $\Omega^{(0)}_{\vec k}$ as given in \eqref{om0eta}.
Given that we have found out in the previous sections that, for several reasons,
the solution of the full equation exhibits unphysical behavior when we keep the
imaginary terms in $\widetilde\omega^2_{\vec k}$, we restrict ourselves here to just considering the case where the imaginary terms in $\widetilde\omega^2_{\vec k}$ are dropped, which leads to the expression \eqref{realomtilde} for the frequency.
Note that we must not remove the imaginary part of the $\Omega^{(0)}_{\vec k}$ appearing as a factor in the first term of the right-hand side of \eqref{dglom1}, because this term originates from the squaring of $\Omega^{(1)}_{{\vec k}} = \Omega^{(0)}_{\vec k}  + \widetilde\Omega^{(1)}_{\vec k}$ and thus ultimately follows from the first term of the right-hand side of \eqref{corrSchr}, which is why it being complex does not violate unitarity.

Therefore, the differential equation we have to solve explicitly reads
\be \label{linom1dgl}
{\rm i}\, \widetilde\Omega^{(1)\prime}_{\vec k}=\frac{2k^3\eta^3 + 2{\rm i}}{\eta(1+k^2\eta^2)}\,\widetilde\Omega^{(1)}_{\vec k} + \frac{H_0^2\eta^4}{2 m_\mathrm{P}^2}\,\frac{k^3(11-k^2\eta^2)}{(1+k^2\eta^2)^3}\,.
\ee
Its general solution is given by
\begin{align} \label{om1linfull}
\widetilde\Omega^{(1)}_{\vec{k}}=&-\,\frac{\eta ^2 \,\E^{-2 \I \eta  k}}{
  \left(\eta k + \I\right)^2} \Biggl\{c_1+\frac{H_0^2}{4 m_\text{P}^2} \biggl[ 9 \E^{-2}\Gamma (0,-2 \I k \eta -2)
\nonumber\\&+3\E^2 \Gamma (0,2-2 \I k \eta)
- \E^{2 \I \eta  k} \frac{(1+\eta  k (\eta  k+6 \I))}{(\eta k-\I)^2}\biggr]\Biggr\},
\end{align}
where $c_1$ is the integration constant. Note that, for convenience, we use the upper incomplete gamma function $\Gamma(0,z)$
instead of its closely related function exponential integral which is sometimes presented as a solution to \eqref{linom1dgl} by computer algebra programs.

In the limit $\eta \rightarrow 0^-$, the Taylor expansion of the function leads to the following form,
\be \label{om1linlimit0}
\widetilde\Omega^{(1)}_{\vec{k}} \approx \eta ^2 \biggl\{c_1+ \frac{H_0^2}{4 m_\text{P}^2} \left(1 + 3 \E^2 \Gamma (0,2)+\frac{9 \Gamma (0,-2)}{\E^2}\right)\biggr\}.
\ee
We see that in the late-time limit, $\widetilde\Omega^{(1)}_{\vec{k}}$ tends to zero regardless of the value chosen for $c_1$.
This integration constant must be fixed by using the Bunch--Davies-like initial conditions obtained in the previous subsection. For such a purpose, we expand our solution for $\eta \rightarrow -\infty$, which leads to
\be
\widetilde\Omega^{(1)}_{\vec{k}} \approx \frac{H_0^2}{4 k^2 m_\text{P}^2}-\frac{c_1 \E^{-2 \I \eta  k}}{k^2}\,.
\ee
From there, it is straightforward to see that one should set $c_1 = 0$ in order to select the only non-oscillatory solution.

Let us mention that there is an extremely good coincidence between the non-oscillating numerical solution
presented in the previous subsection with the solution \eqref{om1linfull} of the linearized equation (with $c_1=0$).
We do not display here the plots corresponding to this latter solution, since they look exactly the same as those presented in Figs.~\ref{plotom01lin} and \ref{plottildeom1lin}. In this latter plot, we also show the Taylor expansion \eqref{om1linlimit0} of the linearized solution around the
origin $\eta=0$. Therefore, we are quite confident that the analytical results obtained here are a very good approximation
to what one would obtain with the solution of the full equation.

In order to calculate the quantum-gravitationally corrected power spectra, we need to insert the real part of the large-scale limit \eqref{om1linlimit0} together with \eqref{om0shor} into \eqref{PS0omega} for the scalar modes,
\begin{align} \label{PS1omega}
{\cal P}_{\text{S}}^{(1)}(k) &=  \frac{4\pi G}{a^2\,\epsilon}\,\frac{k^3}{4\pi^2}\,\left(\Real\Omega_{\vec{k}}^{(0)} + \Real\widetilde\Omega_{\vec{k}}^{(1)}\right)^{-1} \notag\\
&= {\cal P}_{\text{S}}^{(0)}(k)\Biggl[1 - \frac{\Real\widetilde\Omega_{\vec{k}}^{(1)}}{\Real\Omega_{\vec{k}}^{(0)}} + \mathcal{O}\!\left(\frac{H_0^4}{m_{\rm P}^4}\right)\Biggr],
\end{align}
and into \eqref{PT0omega} for the tensor perturbations,
\begin{align} \label{PT1omega}
{\cal P}_{\text{T}}^{(1)}(k) &=  \frac{64\pi G}{a^2}\,\frac{k^3}{4\pi^2}\,\left(\Real\Omega_{\vec{k}}^{(0)} + \Real\widetilde\Omega_{\vec{k}}^{(1)}\right)^{-1} \notag\\
&= {\cal P}_{\text{T}}^{(0)}(k)\Biggl[1 - \frac{\Real\widetilde\Omega_{\vec{k}}^{(1)}}{\Real\Omega_{\vec{k}}^{(0)}} + \mathcal{O}\!\left(\frac{H_0^4}{m_{\rm P}^4}\right)\Biggr].
\end{align}
The real part of \eqref{om1linlimit0} can then be approximated as
\begin{align}
\Real\widetilde\Omega^{(1)}_{\vec{k}} &\approx \frac{H_0^2\eta^2}{4 m_\text{P}^2} \left(1 + 3 \mathrm{e}^2 \,\Real\Gamma (0,2)+\frac{9 \,\Real\Gamma (0,-2)}{\mathrm{e}^2}\right) \notag\\
&\approx -0.988\,\frac{H_0^2\eta^2}{m_\text{P}^2}\,.
\end{align}
Given that $\Real\Omega_{\vec{k}}^{(0)} = k^3\eta^2$ in the large-scale limit, the ratio $\Real\widetilde\Omega_{\vec{k}}^{(1)}/\Real\Omega_{\vec{k}}^{(0)}$ therefore does not exhibit any $\eta$-dependence and thus we have a ``freezing" of the quantum-gravitationally corrected power spectra like in the uncorrected case. However, the $k^3$-term remains, therefore the corrected power spectra become explicitly scale-dependent and we find an \textit{enhancement} of power on the largest scales:
\be\nonumber
{\cal P}_{\text{S},\text{T}}^{(1)}(k) = {\cal P}_{\text{S},\text{T}}^{(0)}(k)\left[1 + 0.988\,\frac{H_0^2}{m_{\rm P}^2}\left(\frac{k_0}{k}\right)^3 + \mathcal{O}\!\left(\frac{H_0^4}{m_{\rm P}^4}\right)\right].
\ee
At this point we have also introduced a reference wave number $k_0$, which represents the inverse of the length scale we have used in the replacement \eqref{repl2}. As we have already mentioned in section \ref{sectpert}, we need to introduce such a length scale in order to make $k$ dimensionful again, such that we can relate it to observable quantities.

From a physical point of view, for the scalar perturbations only the pivot scale used in the analysis of the CMB anisotropies makes sense to be chosen as a reference
scale, because this is the scale at which the scalar amplitude is fixed by observation.

The magnitude of the effect can be estimated by the fact that the tensor-to-scalar ratio of the inflationary perturbations is related to the energy scale of inflation (cf.~e.g.~\cite{Baumann})
\be
\mathcal{V}^{1/4} \sim \left(\frac{r}{0.01}\right)^{1/4}\,10^{16}\,\text{GeV}\,.
\ee
Using \eqref{Vmp}, we find
\be
\frac{H_0^2}{m_{\rm P}^2} = \frac{2\,\mathcal{V}}{m_{\rm P}^4} \sim \frac{2\,r}{0.01}\left(\frac{10^{16}\,\text{GeV}}{m_{\rm P}}\right)^4 \,.
\ee
Given that with our definition of $m_{\rm P}$ \eqref{mpdef}, we have
\be
m_\text{P} = \sqrt{\frac{3}{4\pi G}} \approx 5.97\times10^{18}\,\text{GeV}\,,
\ee
and using the result from the measurements of the Planck satellite \cite{Planck} that $r \lesssim 0.11$, we obtain as an upper limit
\be
\frac{H_0^2}{m_{\rm P}^2} \lesssim 1.74 \times 10^{-10}\,,
\ee
which then immediately leads to the upper limit for the magnitude of the quantum-gravitational correction for $k \sim k_0$:
\be
\left|\frac{{\cal P}_{\text{S},\text{T}}^{(1)}(k) - {\cal P}_{\text{S},\text{T}}^{(0)}(k)}{{\cal P}_{\text{S},\text{T}}^{(0)}(k)}\right|_{k\sim k_0}  \lesssim 1.72 \times 10^{-10}\,.
\ee
We shall discuss the issue of the observability of the corrections in more detail in the follow-up paper.

In the de Sitter case,  the tensor-to-scalar ratio is not influenced by quantum-gravitational corrections due to the fact that the scalar and tensor perturbations are modified by exactly the same correction term, 
\be
r^{(1)} = \frac{{\cal P}_{\text{T}}^{(1)}(k)}{{\cal P}_{\text{S}}^{(1)}(k)} = \frac{{\cal P}_{\text{T}}^{(0)}(k)}{{\cal P}_{\text{S}}^{(0)}(k)} =  r^{(0)} = 16\, \epsilon\,.
\ee
This will no longer hold for a more realistic slow-roll model, to which we shall turn in the next paper.

\section{Conclusions and outlook}

Making use of a semiclassical Born--Oppenheimer type of approximation to the Wheeler--DeWitt
equation, quantum-gravitational effects on scalar and tensor perturbations during inflation have
been computed.
We have found that these effects, described by correction terms to the usual Schr\"odinger equation for the perturbative modes,
lead to an {\em enhancement of power} on the largest scales for both scalar and tensor perturbations.

Let us briefly compare the present results with the results found earlier.
In Ref.~\cite{KK-PRL}, only perturbations of the scalar field were
considered, whereas the perturbations of the metric were neglected.
There a suppression of power was found. Nonetheless, in Ref.~\cite{BEKKP} 
it was pointed out that the considered differential equation,
with the imposed boundary condition, actually allows for two solutions,
one of them becoming discontinuous in some limit. Choosing the continuous solution,
one finds, in fact, an enhancement of power.

We now outline the main differences of the present analysis as compared to \cite{BEKKP}.
First of all, we have incorporated tensor perturbations in our discussion and have dealt with the perturbative gauge invariance for the scalar sector
by constructing the corresponding gauge-invariant master variable, that is, the Mukhanov--Sasaki variable. This allows us to perform
a reduced phase space quantization.
Besides, we have used a different method to implement the Gaussian form of the corrected wave function.
Additionally, we have not a priori neglected the explicitly imaginary terms in the quantum-gravitationally
corrected Schr\"odinger equation \eqref{corrSchr}. In fact, the full solution of this equation has
been numerically analyzed and several physically non-desirable properties produced by
the imaginary terms have been found. Furthermore, we have discussed the natural initial conditions to be imposed
when solving the corrected Schr\"odinger equation, and have defined a modified Bunch--Davies-like
initial quantum state. Finally, this initial state has been used to obtain the numerical solution of \eqref{corrSchr} with
all unitarity-violating terms removed, as well as the analytic solution of the linearized equation \eqref{linearizedeqomega} that gave us the explicit correction to the power spectra.

An important point is the use of the Mukhanov--Sasaki variable $v$ as defined
in \eqref{Mukhdef}. This variable invokes, in particular, a re-scaling with respect
to the scale factor $a$. Such a re-scaling has no effect in the classical theory, but
it is relevant in quantum cosmology where $a$ is quantized. The quantum theory is, in fact,
not invariant with respect to field re-definitions. The use of $v$ is a preferred one,
among other reasons, because it allows a physical regularization of decoherence
factors \cite{BKKM99}. We thus believe that this is the correct variable to use also
in the present context.

In the alternative method to implement the Born--Oppenheimer approximation to the Wheeler--DeWitt equation that was advocated in 
\cite{Sasha1,Sasha2,Sasha3}, an enhancement of power was found for the de Sitter case \cite{Sasha1}, whereas the analysis of the slow-roll case led to a loss of power for the scalar modes, while the tensor modes remained enhanced \cite{Sasha2}. In Ref.~\cite{Sasha3}, the authors adopted the point of view that they can fit the {\em observed} loss of power by their results if they take
the reference scale to be much smaller than the pivot scale; they give a value of
$k_0$ somewhat smaller than 2 Mpc, which corresponds to the scale of galaxy clusters. 
Whether such a small scale turns out to become relevant in quantum gravity, is so far an open, yet intriguing,
question. We note that in all the above-cited papers, including the present one,
the quantum-gravitational correction term is proportional to $k^{-3}$ and thus violates
the (approximate) scale invariance. Such a $k^{-3}$-dependence was already found in \cite{ACT06}.

In loop quantum cosmology, there are several approaches available to
discuss quantum-gravitational corrections \cite{lqc}. In some of them,
also an enhancement of power was found; 
a summarized discussion is given, for instance, in \cite{Gianluca,lqcsum}.
In general, loop quantum cosmology models contain more free parameters
that need to be constrained than the present approach.
The method that most resembles ours is the one outlined in \cite{GMM15,Guillermo15},
and concrete calculations leading to a suppression of power on large scales for scalar perturbations can be found in \cite{BO16}.
Apart from loop quantum cosmology, an
application of supersymmetric quantum cosmology to such a situation
would also be of interest \cite{KLM}. 

In our follow-up paper, we shall apply the methods presented here to the more realistic slow-roll case, which will allow us to discuss the issue of observability more precisely. It can, however, already be said that an effect of the order of $10^{-10}$ on large scales cannot be seen in the CMB anisotropies, because of the statistical uncertainty implied by cosmic variance, which is also most prominent on large scales. It remains to be seen whether the magnitude of the correction turns out to be larger in some alternative, but still observationally justified, inflationary model.

\section*{Acknowledgments}

We thank I\~naki Garay, Alexander Kamenshchik, Christian Steinwachs and Vincent Vennin for fruitful discussions and Alessandro Tronconi for a useful remark on the manuscript. D.\,B.~gratefully acknowledges financial support from the Alexander von
Humboldt Foundation through a postdoctoral fellowship,
as well as from Project IT592-13 of the Basque Government,
and Projects FIS2012-34379 and FIS2014-57956-P of the
Spanish Ministry of Economy and Competitiveness. D.\,B.~also thanks the Institute of Physics of
the University of Szczecin for hospitality while part of this research was done.
C.\,K.~is grateful to the
Max Planck Institute for Gravitational Physics (Albert Einstein
Institute) for kind hospitality while part of this work was done.
The research of M.\,K.~was financed by the Polish National
Science Center Grant DEC-2012/06/A/ST2/00395.
D.\,B.~and M.\,K.~thank the Institute for Theoretical Physics of the University of Cologne, where this work was begun.

\appendix*

\section{Derivation of the final form of the corrected Schr\"odinger equation}
Here, we are going to derive the final form of the quantum-gravitationally corrected Schr\"odinger equation \eqref{corrSchr} using only the minisuperspace variable $\alpha$ and thus neglecting $\tilde\phi$. We start with \eqref{corrSG_raw1}, which restricted to $\alpha$ reads
\begin{align} \label{corrSG_raw2}
\I\,\del{}{\eta}\,\psi^{(1)}_{\vec{k}}= \mathcal{H}_{\vec{k}}\psi^{(1)}_{\vec{k}}+\frac{\E^{-2\alpha}\,\psi^{(1)}_{\vec{k}}}{m_{\rm P}^{2}\,\psi^{(0)}_{\vec{k}}}\Biggl(-\,\frac{1}{\gamma}\,&\del{\psi^{(0)}_{\vec{k}}}{\alpha}\,\del{\gamma}{\alpha} \;\;\\
&+ \frac{1}{2}\,\del{^2\psi^{(0)}_{\vec{k}}}{\alpha^2}\Biggr)\,. \notag
\end{align}
The Hamilton--Jacobi equation of the background \eqref{HJeq} consequently takes the following form
\be \label{HJeqV2}
\left(\frac{\partial S_0}{\partial\alpha}\right)^2 = \E^{2\alpha}\,V(\alpha) \,,
\ee
such that we can write the derivative of the conformal WKB time \eqref{ctimedef} as
\be \label{dereta2}
\del{}{\eta} = -\,\E^{-2\alpha}\,\del{S_0}{\alpha}\,\del{}{\alpha} = -\,\E^{-\alpha}\,\sqrt{V}\,\del{}{\alpha}\,.
\ee
Now, we use this relation and the Schr\"odinger equation \eqref{eq:Schreq} to rewrite the derivative of $\psi^{(0)}_{\vec{k}}$ with respect to $\alpha$,
\begin{align} \label{1derpsi}
\del{\psi^{(0)}_{\vec{k}}}{\alpha}
= \I\,\E^{2\alpha}\left(\del{S_0}{\alpha}\right)^{-1} \I\,\del{}{\eta}\,\psi^{(0)}_{\vec{k}}
=  \frac{\I\,\E^{\alpha}}{\sqrt{V}}\, \mathcal{H}_{\vec{k}}\psi^{(0)}_{\vec{k}}\,.
\end{align}
The second derivative can then be expressed as
\begin{align} \label{2derpsi}
\del{^2 \psi^{(0)}_{\vec{k}}}{\alpha^2}
= \frac{\,\I\,\E^{\alpha}}{\sqrt{V}}\,&\mathcal{H}_{\vec{k}}\psi^{(0)}_{\vec{k}} - \frac{\I\,\E^{\alpha}}{2\,V^{3/2}}\,\del{V}{\alpha}\,\mathcal{H}_{\vec{k}}\psi^{(0)}_{\vec{k}} \\
&+ \frac{\I\,\E^{\alpha}}{\sqrt{V}}\,\del{\mathcal{H}_{\vec{k}}}{\alpha}\,\psi^{(0)}_{\vec{k}} - \frac{\E^{2\alpha}}{V}\left(\mathcal{H}_{\vec{k}}\right)^2\psi^{(0)}_{\vec{k}} \,. \notag
\end{align}
Furthermore, we have to use the condition imposed on the WKB prefactor $\gamma$ in \eqref{gammacond1}, which reads
\begin{align} 
0 &= \frac{1}{\gamma}\,\del{S_0}{\alpha}\,\del{\gamma}{\alpha} - \frac{1}{2}\,\del{^2 S_0}{\alpha^2} \notag\\
&= \E^{\alpha}\left(\frac{\sqrt{V}}{\gamma}\,\del{\gamma}{\alpha} - \frac{\sqrt{V}}{2} - \frac{1}{4\sqrt{V}}\,\del{V}{\alpha}\right),
\end{align}
and can thus be simplified to
\be \label{gammacond2}
\frac{1}{\gamma}\,\del{\gamma}{\alpha} = \frac{1}{2} + \frac{1}{4\,V}\,\del{V}{\alpha}\,.
\ee
Plugging the expressions \eqref{1derpsi} and \eqref{2derpsi} into \eqref{corrSG_raw2} and also using \eqref{gammacond2}, we obtain
\begin{align}
\I\,\del{}{\eta}\,\psi^{(1)}_{\vec{k}} &=\mathcal{H}_{\vec{k}}\psi^{(1)}_{\vec{k}} +\frac{\E^{-2\alpha}\,\psi^{(1)}_{\vec{k}}}{m_{\rm P}^{2}\,\psi^{(0)}_{\vec{k}}}\Biggl[-\,\frac{\I\,\E^{\alpha}}{2\,V^{3/2}}\,\del{V}{\alpha}\,\mathcal{H}_{\vec{k}}\psi^{(0)}_{\vec{k}} \notag\\
&+ \frac{\I\,\E^{\alpha}}{2\sqrt{V}}\,\del{\mathcal{H}_{\vec{k}}}{\alpha}\,\psi^{(0)}_{\vec{k}} - \frac{\E^{2\alpha}}{2\,V}\left(\mathcal{H}_{\vec{k}}\right)^2\psi^{(0)}_{\vec{k}}\Biggr].
\end{align}
Combining the first two terms inside the bracket and using \eqref{dereta2}, we finally recover \eqref{corrSchr}:
\begin{align}
\I\,\frac{\partial}{\partial \eta}\,\psi^{(1)}_{\vec{k}} = \mathcal{H}_{\vec{k}}\psi^{(1)}_{\vec{k}} -\frac{\psi^{(1)}_{\vec{\vec{k}}}}{2\,m_{\rm P}^2  \,\psi^{(0)}_{\vec{k}}}&\Biggl[\frac{\bigl(
\mathcal{H}_{\vec{k}}\bigr)^2}{V}\,\psi^{(0)}_{\vec{k}} \\
&+ \I\,\frac{\partial}{\partial \eta}\!
\left(\frac{\mathcal{H}_{\vec{\vec{k}}}}{V}\right)
\psi^{(0)}_{\vec{k}}\Biggr]. \nonumber
\end{align}

%%%%%%%%%%%%%%%%%%%%%%%%%%%%%%%%%%%%%%%%%%%%%%%%%%%%%%%%%%%%%%%%%%%%%%%%

%%%%%%%%%%%%%%%%%%%%%%%%%%%%%%%%%%%%%%%%%%%%%%%%%%%%%%%%%%%%%%%%%%%%%%%%

\end{document}